\documentstyle[aaspp4]{article}
\def\arcs{\char'175\ ~}
\def\arcsc{\char'175 }

\begin{document}
\title{Dwarf Elliptical Galaxies in the M81 Group: The Structure and
Stellar Populations of BK5N and F8D1\footnote{Based on observations
with the NASA/ESA {\it Hubble Space Telescope}, obtained at the Space
Telescope Science Institute, which is operated by the Association of
Universities for Research in Astronomy, Inc., (AURA), under NASA
Contract NAS 5-26555.}$^,$\footnote{ Observations made with the Burrell
Schmidt of the Warner and Swasey Observatory, Case Western Reserve
University.}}

\author{Nelson Caldwell\footnote{Visiting Astronomer, Kitt Peak
National Observatory, National Optical Astronomy Observatories.}}
\affil{F.\ L.\ Whipple Observatory, Smithsonian Institution, P.O.\ Box 97,
Amado, Arizona 85645}
\affil{caldwell@flwo99.sao.arizona.edu}

\author{Taft E.\ Armandroff}
\affil{Kitt Peak National Observatory, National Optical Astronomy 
Observatories\footnote{The National Optical Astronomy Observatories
are operated by AURA, Inc., under cooperative agreement with the National
Science Foundation.},\\
P.O.\ Box 26732, Tucson, Arizona 85726}
\affil{armand@noao.edu}

\author{G. S. Da Costa}
\affil{Mount Stromlo and Siding Spring Observatories, The Australian National
University, Private Bag, Weston Post Office, ACT 2611, Australia}
\affil{gdc@mso.anu.edu.au}

\author{Patrick Seitzer}
\affil{Department of Astronomy, University of Michigan, Ann Arbor, Michigan
48109}
\affil{seitzer@astro.lsa.umich.edu}

\begin{abstract}

The M81 group is one of the nearest groups of galaxies, but its
properties are quite different from those of the Local Group.  It has
therefore provided a different environment for the evolution of its member
galaxies.  We have carried out a CCD survey of the M81 group to search
for analogs to Local Group dwarf elliptical (dE) galaxies.  All the M81
dwarfs previously identified in photographic surveys were recovered and
we also discovered several new systems whose surface brightnesses fall within
the range found for Local Group dE's.  We have obtained HST WFPC2 images
through the F555W and F814W filters of two M81 group dE's: BK5N and a new
system, designated F8D1.  The resulting color--magnitude diagrams show
the upper two magnitudes of the red giant branch.  The I magnitudes of the
red giant branch tip in both galaxies yield distances that are consistent
with membership in the M81 group.  Surface brightness and total
magnitude measurements indicate that BK5N and F8D1 have similar
central surface brightness (24.5 and 25.4 mag/arcsec$^2$ in V,
respectively), but F8D1's larger length scale results in it being 3
magnitudes more luminous than BK5N.
BK5N lies on the relation between central surface brightness and
absolute magnitude defined by Local Group dwarf ellipticals, but F8D1
does not.  F8D1 is more luminous for its central surface brightness
than the relation predicts, similar to the large low surface
brightness dwarf galaxies found in, for example, Virgo.  The mean color of
the giant branch is used to establish the mean abundance of each galaxy.
F8D1, the more luminous galaxy, is significantly more metal rich ([Fe/H]
$\approx$ --1.0) than BK5N ([Fe/H] $\approx$ --1.7).  Both BK5N and F8D1 lie
on the relation between absolute magnitude and metal abundance defined by
Local Group dwarf ellipticals.  However, as regards the relation between
central surface brightness and metal abundance, BK5N again follows the Local
Group dwarfs, while F8D1 deviates significantly from this relation.
This suggests that the total amount of luminous matter is more fundamental
in controlling metal enrichment than the surface density of luminous
matter.  We have also used the color width of the giant branch compared
with the photometric errors to establish abundance ranges in both
galaxies, the sizes of which are comparable to those in Local Group dE's.
From the numbers and luminosities of asymptotic
giant branch stars more luminous than the red giant branch tip, we infer
that, again like many of the Local Group dE's, both BK5N and F8D1 have
had extended epochs of star formation.  F8D1 contains stars as young as
3 -- 4 Gyr, while in BK5N stars as young as $\sim$8 Gyr are present.  The
fractions of intermediate-age population, at $\sim$30\%, are similar in both
galaxies.  Further, one globular cluster has been found in F8D1, but none
are present in BK5N.  These numbers of clusters are consistent with our
expectations from Local Group dwarfs.  Overall, we find that BK5N is
similar in all respects to the Local Group dE's.  Thus, in spite of the
different environments of the M81 group and the Local Group, dwarf
galaxies of very similar properties are present in each group.  F8D1, on
the other hand, has no Local Group counterpart and is indeed rare
even in large clusters of galaxies. Our study of its
stellar population is the first for a large low surface brightness
dwarf.
\end{abstract}

\section{Introduction}

Within the Local Group the most numerous type of galaxy are the dwarf
elliptical\footnote{We consider the dwarf spheroidals to be a subset of
the dwarf ellipticals, with the former term referring specifically to
the lowest luminosity gas-poor galaxies found around the Galaxy, and
the latter term encompassing all low surface brightness, gas-poor
galaxies regardless of location or luminosity.} (dE) systems: there
are now nine known dE companions to the Milky Way, six dE
companions to M31 and one apparently isolated dE, Tucana.  Because of
their proximity, these galaxies are the most easily studied examples of
what may be the most common type of galaxy in the Universe.  In the
last decade these local systems have been studied in increasing detail
and though their evolutionary history is far from completely
understood, these dE systems are all now relatively well observed
objects (see, for example, recent reviews by Da Costa 1992, Zinn 1993
\& Armandroff 1994).  Indeed the Milky Way companion dE galaxies show
well established relationships between surface brightness, absolute
magnitude, length scale and mean abundance which both the M31 dE
companions and the Tucana system also follow.  This would seem to
suggest that these relations are characteristic of the dE formation
process and that this process is not strongly dependent on the
environment in which the dE systems are found.  On the other hand,
there is some evidence that these relations are not universal.  For
example, Caldwell et al.\ (1992) have included the low surface
brightness Virgo cluster dE galaxies from Impey et al.\ (1988) in a
plot of central surface brightness against absolute magnitude for the
Local Group dE systems.  They conclude (cf.\ Caldwell \&
Bothun 1987; Impey et al.\ 1988) from this diagram, with its larger
sample of dE galaxies, that the relationship between surface
brightness and absolute magnitude breaks down at faint surface
brightness ($V\ge24$): below this level dE galaxies can be found with a
wide range of luminosities.  This immediately raises the question: are
the abundances and stellar populations of such low surface brightness
systems determined by their total luminosities or by their central surface
brightnesses?  Support for either alternative can be found in
theoretical models in the literature (e.g.\ Phillipps et al.\ 1990,
Bender et al.\ 1993).  But by studying the properties of dE galaxies
outside the Local Group, we can begin to answer this question
observationally and thus address the broader issue of the extent to
which environment affects the properties of these systems.

The M81 group of galaxies is one of the nearest groups to our own.  It
contains one large spiral (M81), two peculiar galaxies (M82 and NGC
3077), a few small spirals and a large number of dwarf galaxies.
Further, there is evidence that a number of substantial interactions
have occurred in this group in the recent past (Solinger et al.\ 1977;
Kennicutt et al.\ 1987; Yun et al.\ 1994).  Thus it provides a quite 
different environment from that of the Local Group, enabling the 
possibility of comparative studies.

In this paper, we discuss the structure and stellar populations of two
low surface brightness galaxies in the M81 group.  We begin, in the
next section, by summarizing a CCD survey for low surface brightness
galaxies in the vicinity of M81 and describing how we selected two
galaxies for detailed study.  Sec.\ \ref{photometry} describes our
HST/WFPC2 imaging of these two galaxies and the resulting photometry of
these galaxies' stars.  The color--magnitude diagrams that result from
this photometry are presented in Sec.\ \ref{cmd}.  The color--magnitude
diagrams are analyzed to yield distance moduli in Sec.\ \ref{distance},
metal abundances in Sec.\ \ref{metallicity}, and inferences about the
intermediate-age population based on upper asymptotic giant branch
stars in Sec.\ \ref{uagb}. The globular cluster content of these
galaxies is discussed in Sec.\ \ref{gc}.  Finally in
Sec.\ \ref{discussion}, we compare these two M81 group dwarfs with their
Local Group counterparts, concentrating our analysis on the central
surface brightness--absolute magnitude diagram and the correlations
between metal abundance and both absolute magnitude and central surface
brightness.

\section{Selection of the M81 Group Dwarfs}
\subsection{A new CCD survey}

In a number of papers, Borngen and colleagues (Borngen et al.\ 1982a,
1982b, 1984; Karachentseva et al.\ 1984, 1985) have surveyed the M81
group for dwarf galaxies using photographic plates from the Tautenberg
Schmidt telescope, attempting to supplement the catalogs of brighter
galaxies such as those of de Vaucouleurs (1975) and Kraan-Korteweg \&
Tammann (1979).  This work ultimately resulted in the catalog of
Karachentseva et al.\ (1985) which listed 37 galaxies as possible dwarf
members of the group, down to a quoted photographic magnitude of
about 18.  Binggeli (1983; see also van Driel et al.\ 1997) edited this list
somewhat, showing 41 galaxies with apparent photographic magnitudes
from 8 to 18.

While the quoted limiting magnitude of 18 corresponds to M$_{\rm V}=-10$,
thus placing the candidates in the dwarf spheroidal regime, it seemed
to us that a further survey of the group using a CCD as a detector
would better explore that luminosity realm. To this end, we obtained
new images of the M81 field with the Burrell Schmidt on Kitt Peak and a
Tektronix/STIS 2048$\times$2048 CCD, located at the Newtonian focus.
This system provides images with a 1 degree field, with 2.03
arcsec/pixel.  Over the years 1991-1994 we obtained 60 overlapping
V-band frames around the M81/M82/NGC3077 trio, resulting in an areal
coverage of approximately 40 square degrees.  Exposure times were 1500
sec.  Unfortunately, bad weather was a constant companion during the
years of observations. Hence most of the frames were taken during good,
but not photometric, conditions.  As a result, it has proved to be
impossible to provide accurate survey limits for all the fields, though
our newly detected galaxies tell us that we did go deeper than the
photographic surveys.

The areal coverage around M81 is equivalent to a radius of about 180
kpc; this radius around our Galaxy includes all but the two outer
Galactic dwarf ellipticals Leo I \& II.  We easily recovered all of the
previously cataloged M81-group dwarfs, and found seven more candidates
as well.  Of these, one was not confirmed on re-observation with a
different telescope and one turned out to be a Galactic reflection
nebula (common in the M81 group, see Sandage 1976).  Three others have
been confirmed to be dwarf members of the M81 group, all apparently
dwarf ellipticals, while the nature of the final two awaits future
observations.

In this paper, we report further extensive observations of one of the
confirmed new dwarf ellipticals as well as a dwarf elliptical listed in
Karachentseva et al.\ (1985), in fact the faintest one in that
catalog.

\subsection{Ground Based Imaging of M81-BK5N and M81-F8D1}\label {ground}

From the Karachentseva et al.\ (1985) catalog we have chosen
BK5N. A V band CCD image from the FLWO 1.2m telescope is shown in Fig. 
\ref{combined_pic}.  
This galaxy appears to have a nucleus,
off-center by about 2\arcsc. The HST images have revealed this ``nucleus''
in fact to be a background spiral galaxy.

The galaxy from our own survey that we discuss here is named F8D1 (it
was found on our eighth field), also shown in Fig.\ \ref{combined_pic}.
It is evident that this galaxy is of
very low central surface brightness, yet it is also quite large on the sky,
much larger than BK5N.  There is also some apparent asymmetry in the outer
isophotes at the faintest levels, which we will discuss further below.

\begin{figure}[p]
\plotone{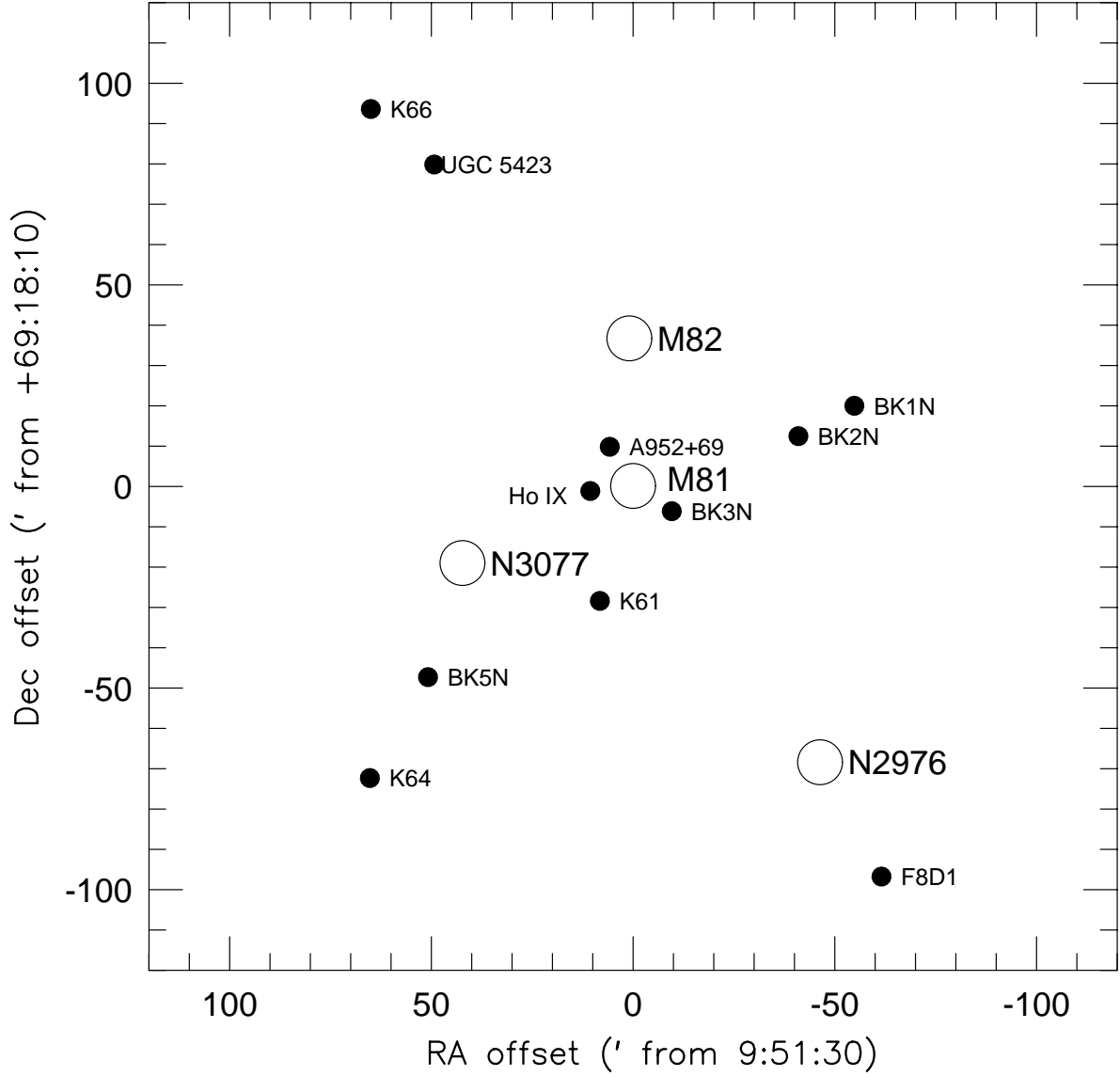}
\caption[]{Positions on the sky of the two galaxies observed here (BK5N
and F8D1)
with respect to other members of the M81 group within a projected radius of
125 kpc from M81 itself (assuming M81 is 3.6 Mpc distant).}
\label{pos_fig}
\end{figure}

Figure  \ref{pos_fig} shows the position of these two galaxies on the sky with
respect to the brighter galaxies in the group.  F8D1 is about 1/2
degree away from the spiral NGC 2976, corresponding to a separation of
34 kpc if both galaxies are at the distance of M81 (3.6 Mpc, Freedman
et al.\ 1994).  In fact, an accurate distance of NGC 2976 is not
available, and that of F8D1 is somewhat larger than that of M81
(see Sec.\ \ref{distance}).  The
implied tidal radius of F8D1 caused by NGC 2976 would be R$_t$ = 6.5
kpc (assuming the same M/L for both galaxies), whereas the actual
limiting radius of F8D1, determined from the surface brightness profile,
is 5.8 kpc. It is thus possible
that F8D1 is tidally limited by NGC 2976, if both galaxies are at the same
distance.  BK5N lies at a similar angular distance from NGC 3077,
implying a tidal radius for the former galaxy of 2.1 kpc.  The dwarf's
actual limiting radius is only 0.9 kpc, so this galaxy might also be
tidally limited by the nearby giant, again assuming similar distances
(there is no accurate distance to NGC 3077 either).
But it is more probable that the radii of both galaxies were determined during
formation, because there are other dE's with similar light profiles in
clusters that are not near large galaxies (Caldwell \& Bothun 1987).

Our calculations showed that giant branch tip stars in galaxies at the
distance of M81 should be resolvable from the ground in excellent
seeing.  Because of its crude resemblance to the surrounding Galactic
cirrus, F8D1 was imaged with a CCD on the MDM 2.4m telescope in the R
band during two runs in 1994 and 1997, in an attempt to verify its
nature as a galaxy in the M81 group before proceeding with HST
observations.  The 1994 data had 0.7\arcs seeing, while the 1997 data
had 1.2\arcs seeing.  Figure \ref{mdm_pic}, a mosaic of both CCD
frames, reveals that indeed the galaxy is well resolved into stars,
including the faint asymmetric outer part to the north.  Thus the
asymmetry is not due to contaminating cirrus but is rather a feature of
F8D1; however it is a relatively minor contributor to the total
luminosity of the galaxy, only a few percent.  It is unlikely that the
feature is a tide raised by NGC 2976, because tides raised on dwarf
galaxies by giants are expected to be symmetric (Piatek \& Pryor
1995).  Whatever the cause, other small asymmetries are observed in
some of the Local Group dE galaxies (e.g.\ Irwin \& Hatzidimitriou
1995).

\begin{figure}[p]
\plotone{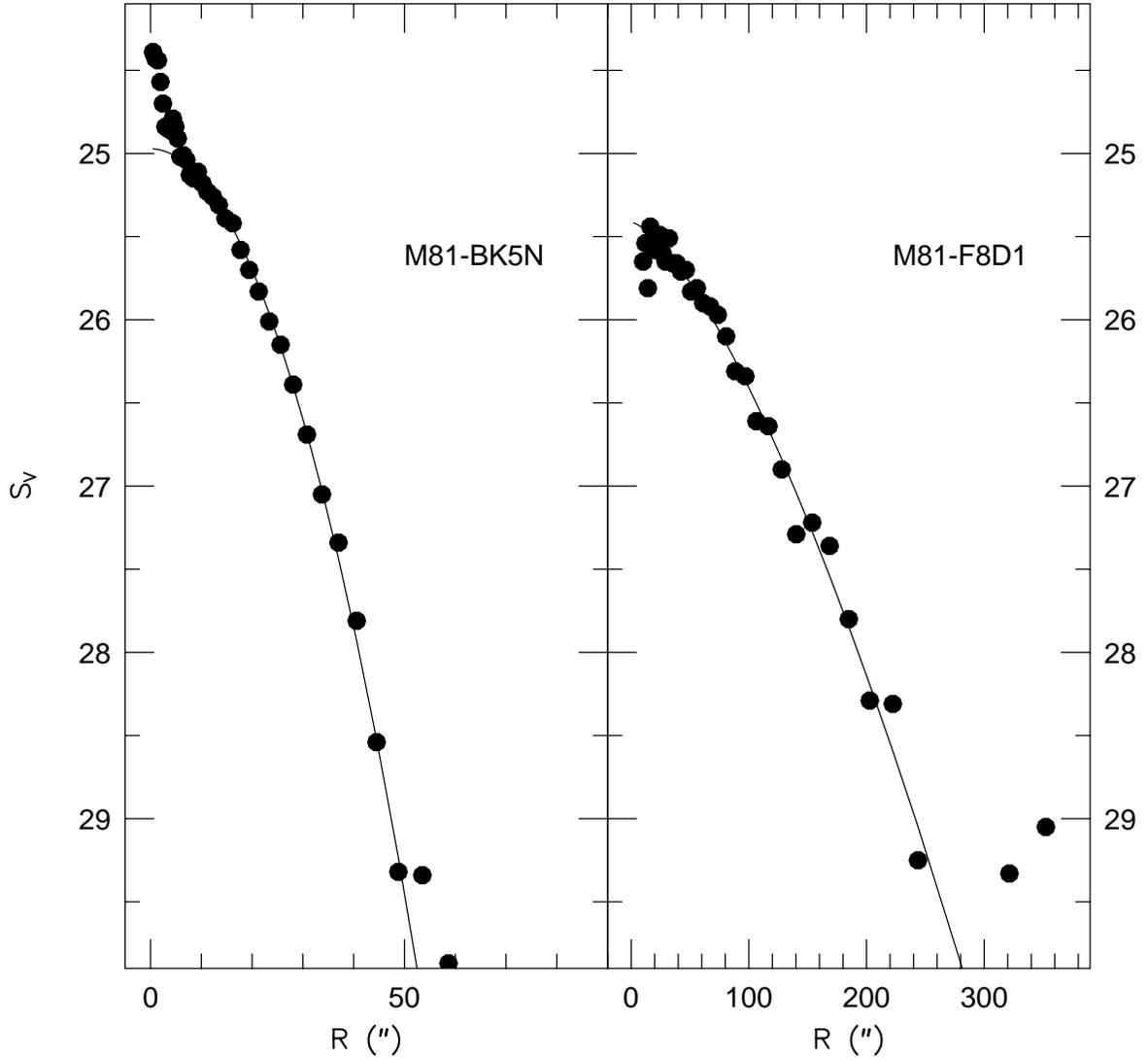}
\caption[]{Radial V light  profiles of BK5N and F8D1.  Surface brightness 
is plotted in $V$ magnitude per square arcsec against the geometric
mean of the major and minor axes of the best fitting elliptical
isophotes, in arcsecs.  Note the scale change in
horizontal axis between BK5N and F8D1.  The continuous line in each
panel is the best fit Sersic profile, using the parameters in 
Table \ref{phot_table}.}
\label{profile_fig}
\end{figure}

BV CCD images from the Burrell Schmidt and the FLWO 1.2m telescope were
used to derive the basic photometric properties and light profiles of
these two galaxies.  The techniques described in Caldwell et
al.\ (1992) were employed in the analysis of these images.  Photometric
zero points were derived from standard stars observed during the nights
of observations.  The ellipticity of F8D1 was assumed to be 0.0, as
judged by eye, while that of BK5N was measured via an ellipse fitting
program to be 0.4. The asymmetry of F8D1 was ignored in the photometry
because it contributes only a few percent to the total luminosity.
Likewise, the central background galaxy in BK5N contributes about 1\%,
and was interpolated across in the image before analysis (which was not
completely successful, as the figure still shows a steepening of the
profile in the center).  Figure \ref{profile_fig} shows the light
profiles, and Table \ref{phot_table} lists the derived properties.  The
profile of F8D1, while exponential in the outer parts, is shallower
than an exponential in the center; the profile of BK5N is much steeper
than an exponential (common for the faintest dE's, see Caldwell \&
Bothun 1987).  Thus Sersic profiles (I=I$_{\rm 0}e^{-(r/r_{\rm 0})^n}$,
Sersic 1968), which have an additional free parameter $n$ over an
exponential (which corresponds to $n$=1), were fit to the data.  The
derived parameters are shown in Table \ref{phot_table}.  In addition to
being much smaller, BK5N has a much steeper profile than F8D1, judging
from the differences in the slope parameter $n$.

A general comparison of the two galaxies reveals F8D1 to be about six
times larger in radius than BK5N, even though the latter galaxy has a
somewhat higher central surface brightness.  Thus F8D1 is much more
luminous than BK5N; it is also slightly redder (Table \ref{phot_table}), 
though the photometric errors are large.
Placing these galaxies in the
surface brightness--luminosity diagram for dE's
(Fig.\ \ref{sb_lumin_fig}) allows some further comments to be made (the
distances used in obtaining the luminosities are derived 
in Sec.\ \ref{distance}).  BK5N
is clearly similar to the Galactic and M31 dwarf ellipticals - it is
small and of low luminosity (it is also highly flattened).  {\it F8D1 on the
other hand has no counterpart in the Local Group}.  Rather, its nearest
analogs can be found in the Virgo and Fornax clusters, which contain a
small number of very large, low surface brightness dwarfs, as described
by Sandage \& Binggeli (1984) and studied in detail by Impey et al.
(1988) and Bothun et al.\ (1991), who showed histograms of scale lengths
and surface brightnesses for Virgo and Fornax cluster dwarf
ellipticals. The scale length of F8D1 at Virgo 
would be about 18-21\arcsc which is comparable to the scale lengths of Virgo
systems V2L8 and V3L1, for example, dE's which have similar central surface 
brightnesses to F8D1 (Impey et al.\ 1988). Nevertheless fewer than 1\% of 
cataloged Virgo or Fornax dE 
galaxies with similarly low central surface brightnesses are as large as F8D1,
less than 10 in number.  Thus F8D1 is a rare type of galaxy, in any
environment.  F8D1 is the first such large scale length dwarf galaxy to
be resolved into stars. For BK5N, the scale length at Virgo would be 3
to 4\arcsc, which is basically below the current detection limit in
those clusters, though as we have noted, it is similar to many of the
Local Group dwarf ellipticals.

Finally, from the ground-based images these two galaxies do not appear
to have had recent star formation.  Huchtmeier \& Skillman (1994) and
van Driel et al.\ (1997) have surveyed previously cataloged M81 group
galaxies at 21 cm, but did not detect any H {\sc i} in BK5N\@.
Assuming a line width of 24 km/sec (corresponding to a typical Galactic
dE velocity dispersion of $\sim$10 km/sec), the 3$\sigma$ upper limit
for BK5N of 13.5 mJy given by van Driel et al.\ (1997) corresponds to
M$_{\rm HI}$/L$_{\rm B}$ $<$ 0.4 in solar units, provided the velocity
of the dwarf is sufficiently separated from that of the Galactic H {\sc
i}.  This is not a very strong limit since Local Group dE's have M$_{\rm
HI}$/L$_{\rm B}$ $\lesssim$ 10$^{-2}$, while the Local Group dwarfs
Phoenix and LGS3, which contain young stars, have H {\sc i} detections
that correspond to M$_{\rm HI}$/L$_{\rm B}$ $\approx$ 0.1 -- 0.2.
Deeper H {\sc i} observations of BK5N, and of F8D1 which has not been
observed in H {\sc i} to our knowledge, are clearly desirable.

\section{HST WFPC2 Photometry} \label{photometry}

To explore the resolved stellar populations of these two galaxies, in
particular to measure the distances, the mean stellar metallicities,
the metallicity dispersions and the young/intermediate age star
content, we observed BK5N and F8D1 with the WFPC2 camera on HST.
Previous work on deriving metallicities of old stellar populations has
indicated that the giant branches must be observed down to about 2 mags
below the tip for greatest accuracy.  If our two dwarfs are at the
distance of M81, this means that stars down to I $\sim$ 26 must be
observed.  This number dictated the exposure times for our galaxies.
We elected to observe with the F555W and the F814W filters, so that we
could transform easily to VI magnitudes, and hence employ the techniques
of Da~Costa \& Armandroff (1990) using the I magnitude of the giant
branch tip for distance determinations and the (V-I)$_0$ color
of the giant branch for metallicity measurements.

Our objects are located in the so-called ``continuous viewing zone'' of
HST, meaning that they could be observed continuously through several
orbits.  Scheduling difficulties prevented us from obtaining all of the
data at a single visit however, hence both galaxies were observed
at two different spacecraft orientations.  This meant that (1) we could
not combine the frames themselves together, rather we would have to
combine the resultant stellar photometry instead and (2) the area
surveyed at the total exposure depth would only be the overlapping
areas of the two visits, rather than the full field of the WFPC2
frames.  On the positive side, the time span between observations
allowed us to search for variable stars.  Table \ref{obs_log_table}
presents the observation log, and Fig.\ \ref{combined_overlay} 
shows overlays of the WFPC2 footprint for the two
visits for each of the galaxies.  Most of BK5N will thus be found on
the PC frame alone, while even the entire WFPC2 area would not include
all of F8D1, let alone the overlapping areas from the two visits.

The frames were debiased, zero-corrected, dark-subtracted and
flat-fielded by the STScI pipeline process
before being made available to us.  We then
combined the frames for each filter and visit using the STSDAS routine
gcombine, which effectively removes cosmic rays.  Since we would be
doing point source photometry, we also needed to restore the
photometric flatness of the images, which had been changed by the
flat-fielding process (which corrects for effect of geometric
distortion of the images on extended objects at the expense of
introducing a non-flatness for point sources).  The frames were multiplied
by the so-called ``geometric-distortion'' frame to accomplish the
correction.

\begin{figure}[p]
\plotfiddle{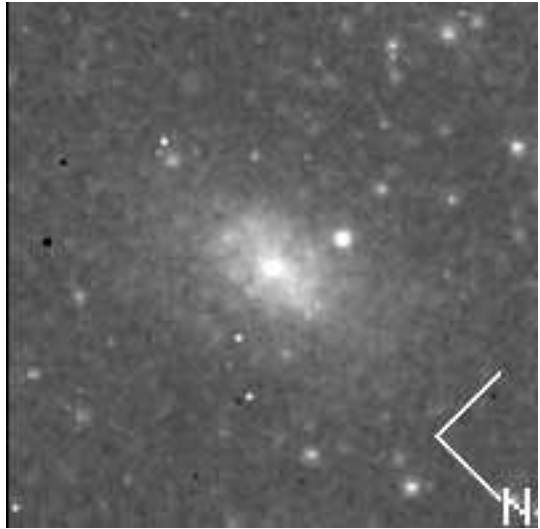}{3.0in}{0.}{100.}{100.}{-301.}{-220.}
\caption[]{Central object of BK5N from the combined F555W PC images
of both epochs.  The direction indicator lines are 0.75\arcs long.
The object appears to be a background galaxy.}
\label{nuc_gal_pic}
\end{figure}

Figure \ref{bk5n_wfpc2} shows the PC field for one of the
visits to BK5N, and Fig.\ \ref{f8d1_wfpc2} shows the full WFPC2 field for
F8D1.  It is clear that both galaxies are well-resolved into stars, and
that stellar photometry would not be compromised by the minimal
crowding.  As mentioned above, BK5N seems to be nucleated from the
ground based images, with the ``nucleus'' 2\arcs offcenter.
(Note that the ``N'' in the name given by Karachentseva et al.\ 1985 refers
to ``new'', not ``nucleated''.)  The PC image shows that this
``nucleus'' is in fact a background spiral galaxy (see enlargement in
Fig.\ \ref{nuc_gal_pic}), with V--I = 1.2, typical for field galaxies
of its magnitude (see Driver et al.\ 1995).

Instrumental magnitudes were measured from the combined frames using
conventional digital aperture photometry, as implemented in the IRAF
package {\sc daophot}.  In particular, the {\sc find} task was used to
identify the stars, and the {\sc phot} task provided the photometry.
An aperture radius of 2.0 pixels was used for both the WF and PC
frames.  The next step was to determine the aperture correction from
the 2-pixel-radius apertures to the standard aperture size for the
WFPC2 photometric system, 0.5 arcsec in radius (Holtzman et
al.\ 1995).  These aperture corrections were measured using fairly
bright, uncrowded stars.  Because only a few suitable stars, at most,
were present on each CCD, we were forced to ignore any spatial
variations in the aperture corrections.  Thus, for each visit and
filter, we adopted a single aperture correction for the WF frames and
another for the PC frame.  As noted in Da Costa et al.\ (1996),
ignoring the spatial variation in the aperture corrections results in
only a second order effect on the colors.  After applying the aperture
corrections, we corrected for exposure time and the gain factors for
each CCD (following Holtzman et al.\ 1995).  We also eliminated stars
found on the vignetted regions of the CCDs (X$<$75, Y$<$75 for the WF
CCDs; X$<$100, Y$<$100 for the PC) and a small number of obvious galaxies.

We derived transformation equations between the coordinate system of the
first visit and that of the second visit for each galaxy and filter.
This allowed us to match the photometry for stars in common between the
two visits.  Because the WFPC2 orientation on the second visit for each
galaxy is rotated significantly from the orientation on the first visit
(see Fig.\ \ref{combined_overlay}), comparing the
magnitudes during the two visits yields information on the
charge-transfer ramp effect (see Holtzman et al.\ 1995) during our
observations.  In plots of magnitude difference between the two visits
versus change in Y coordinate, we could see no evidence of the
charge-transfer ramp.  This may be a consequence of our relatively long
exposure times, and hence relatively high background levels.  In any
case, we decided not to apply any charge transfer ramp correction.
Next, we combined the corrected magnitudes for the two visits weighted
according to their uncertainties and only
retained stars that had photometry on both visits in both
filters.  Finally, we transformed from the instrumental system to
standard $V,I$ using the zeropoints and color terms of Table 7 of
Holtzman et al.\ (1995).

We have compared the uncertainties in the I magnitudes and V--I colors
based on photon statistics with the uncertainties implied by the
repeatability of the magnitudes and colors between the two visits.
Table \ref{error_tab} lists, as a function of I magnitude, the mean
uncertainties based on photon statistics ($<\sigma_{\rm photon}>_I$ and
$<\sigma_{\rm photon}>_{V-I}$) and those based on repeatability between
the two visits ($<\sigma_{\rm repeat}>_I$ and $<\sigma_{\rm
repeat}>_{V-I}$) after excluding variable stars (see
Sec.\ \ref{variables}).  The uncertainties computed by the two methods
agree very well at magnitudes fainter than I $\approx$ 24.5.  Brighter
than this, the repeatability uncertainties do not decrease as rapidly
as those based on the photon statistics, particularly for the I
magnitudes.    This suggests that there are contributions from effects
such as flat fielding, dark subtraction, or the frame combination
process, for example, which limit the errors at the brighter
magnitudes.  Such a result was also noted by Da~Costa et al.\ (1996).
In addition to these photometric errors,
we will also consider in the subsequent analysis the effects of a possible
$\pm$0.05 mag systematic zeropoint uncertainty.  This comes from the uncertainty
in the aperture corrections ($\sim\pm$0.04 mag) and from uncertainty in the
zeropoint of the transformed photometry relative to the standard system
($\pm$0.03 mag).

\section{Color--Magnitude Diagrams} \label{cmd}

\begin{figure}[p]
\plotone{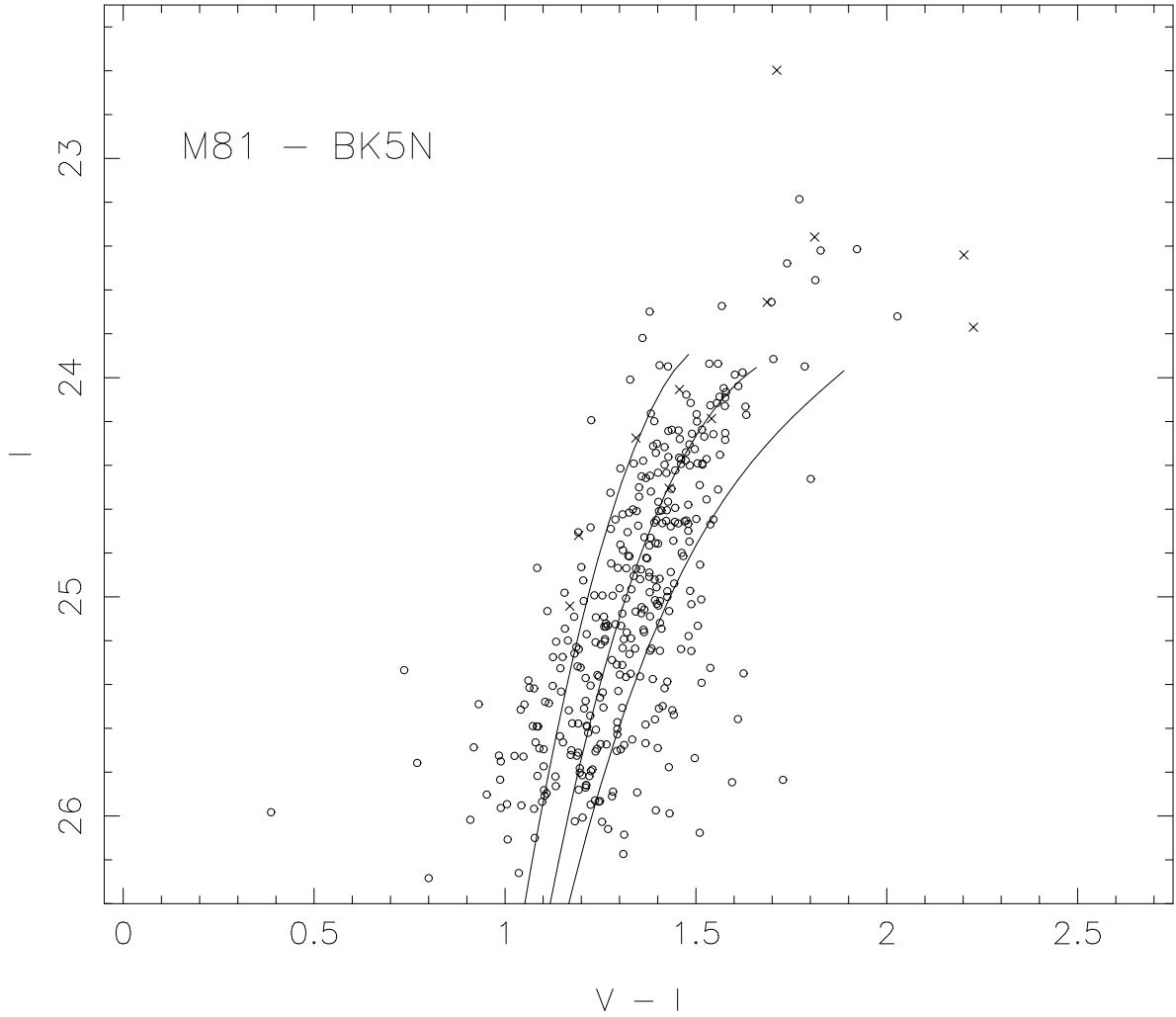}
\caption[]{Color--magnitude diagram for M81 BK5N (PC frame only).
Standard globular cluster giant branches are shown for M15 ([Fe/H] =
--2.17), M2 (--1.58), and NGC 1851 (--1.16) (Da Costa \& Armandroff
1990) shifted to the distance modulus and reddening of BK5N
\mbox{(($m$--$M$)$_0$} = 27.9; E($V$--$I$) = 0.065).  The $\times$
symbols represent stars with significant photometric variations between
the two visits.}
\label{bk5n_cmd}
\end{figure}

\begin{figure}[p]
\plotone{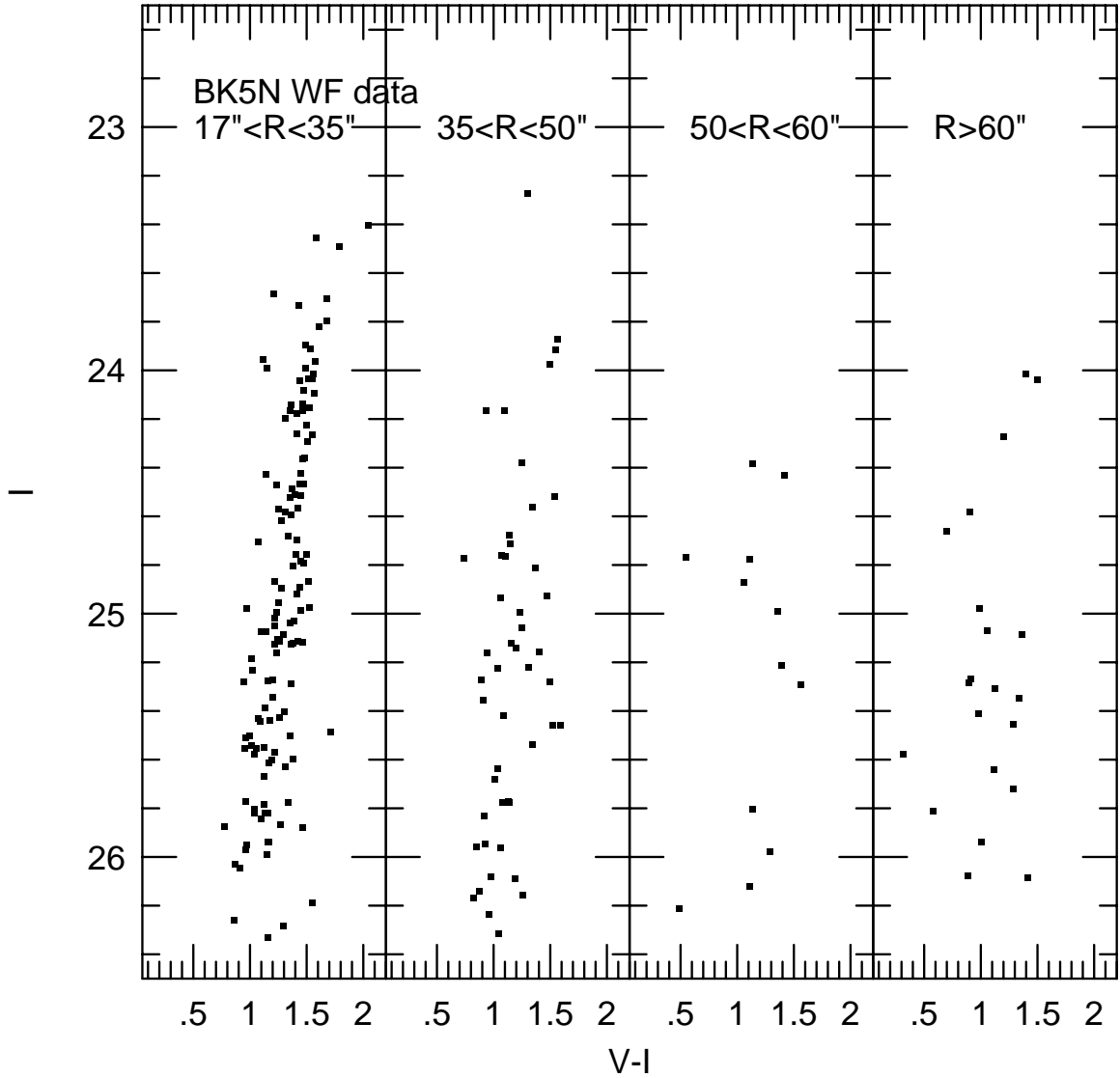}
\caption[]{Color--magnitude diagram from WF data of BK5N.  The four
different panels show diagrams based on stars at increasingly larger
radii from the center of BK5N, in terms of the geometric means of
the semi-axes of the elliptical isophotes found from the ground 
based data.  The limiting radius of BK5N is about 50\arcsc, and indeed
we see few stars that could be BK5N members beyond this distance.
Overall, the field contamination would appear to be
very small for either of our two galaxies.}
\label{field_cmd}
\end{figure}

Figure \ref{bk5n_cmd} shows the I,V--I color--magnitude diagram of BK5N
for the 357 stars measured on the PC frames in both datasets.  As
discussed previously, and as is apparent from
Fig.\ \ref{combined_overlay}, the bulk of BK5N falls on the PC frame,
and the WF frames cover mostly background.  This is further illustrated
in Fig.\ \ref{field_cmd} which shows the color--magnitude diagram of
the BK5N WF frames as a function of distance from the center of BK5N.
In Fig.\ \ref{field_cmd} only the innermost portion of the WF frames
shows a strong indication of a giant branch.  The BK5N PC
color--magnitude diagram in Fig.\ \ref{bk5n_cmd} reveals a well defined
giant branch, with a giant branch tip magnitude of I $\approx$ 23.9.
No bright blue stars are present.  Anticipating the BK5N distance
modulus and mean abundance results from the next two sections, and
using the Bertelli et al.\ (1994) log(Z/Z$_{\sun}$) = --1.3 isochrones,
the lack of stars with V--I $\leq$ 0.7 in Fig.\ \ref{bk5n_cmd}
indicates that no star formation has occurred in this dwarf for at
least the last $\sim$300 Myr.  However, a significant number of red
stars more luminous than the giant branch tip are evident in
Fig.\ \ref{bk5n_cmd}.  These stars have no counterpart in the
background color--magnitude diagram (see the outer two bins in
Fig.\ \ref{field_cmd}) and are most probably upper asymptotic giant
branch (AGB) stars in BK5N\@.  They will be discussed in
Sec.\ \ref{uagb}.

Figure \ref{f8d1_cmd} shows the I,V--I color--magnitude diagram for
F8D1 based on the 2061 stars measured on the WF frames from both
datasets.  Unlike the case for BK5N, the F8D1 WF frames are nicely
populated with member stars (see Figs.\ \ref{combined_overlay} \&
\ref{f8d1_wfpc2}).  As with BK5N, a giant branch is evident in the
color--magnitude (c-m) diagram and the giant branch tip magnitude is I
$\approx$ 24.0 (quite similar to that of BK5N).  Again, stars more
luminous than the giant branch tip are evident, and no blue stars are
present.  Combining the F8D1 distance modulus and mean abundance
results from the next two sections with the Bertelli et al.\ (1994)
isochrones for log(Z/Z$_{\sun}$) = --0.7 then suggests that no star
formation has occurred in F8D1 for at least the last 150 Myr.  Figure
\ref{f8d1_cmd} does, however, show a small (31 stars) population
fainter than the red giant branch tip (by $\gtrsim$ 0.8 mag) with
intermediate colors (0.2 $\lesssim$ V--I $\lesssim$ 1.0).  Scaling the
number of the stars with 24.5 $\leq$ I $\leq$ 25.7 and V--I $\leq$ 1.0
by the appropriate areas then predicts that the BK5N ``field'' region
c-m diagram (R $>$ 60$\arcsec$) should contain 11 similar objects.  In
fact the c-m diagram (Fig.\ \ref{field_cmd}) contains 5 stars in the
same color and magnitude range.  Allowing for Poisson statistics
variations in the actual numbers of stars, this difference is a
$\sim$2$\sigma$ effect.  We thus interpret the minor population of
intermediate color stars in the F8D1 c-m diagram as due primarily to
foreground stars, together with a probable component of F8D1 stars
arising from occasional larger than average photometry errors.  These
stars will not be discussed further.  Overall, the c-m diagrams for
these M81 group dwarf galaxies are thus quite similar to those for
Local Group dwarf ellipticals.

Again in anticipation of the distances to BK5N and F8D1 determined in
Sec.\ \ref{distance}, we have overplotted the color--magnitude diagrams
in Figs.\ \ref{bk5n_cmd} and \ref{f8d1_cmd} with standard globular
cluster giant branches shifted to the distance modulus and reddening
appropriate for each galaxy.  The reddenings for BK5N and F8D1 come
from the Burstein \& Heiles (1982) extinction maps and are listed in
Table \ref{phot_table}.  The standard globular cluster giant branches
are from Da Costa \& Armandroff (1990).  The metal abundances of these
galaxies and whether a range of abundance is present in each system
will be discussed in Sec.\ \ref{metallicity}.  For now, we note that
the mean BK5N giant branch lies near that of M2 (which has [Fe/H] =
-1.58) and the mean F8D1 giant branch lies between the giant branches
of NGC 1851 and 47 Tuc (which have [Fe/H] = -1.16 and -0.71,
respectively).  The abundances for these M81 group systems are thus
also quite similar to those for Local Group dwarfs.

\subsection{Variable stars} \label{variables}

Because of the time separation between the two HST visits, we can look
for variability in our data.  For BK5N, the two epochs of observation
were 109 days apart, while for F8D1 the separation was 56 days. Such a
long time span allows us to search particularly for long period or
irregular variables that are expected to occur among giants near, and
above, the red giant branch tip.

\begin{figure}[p]
\plotfiddle{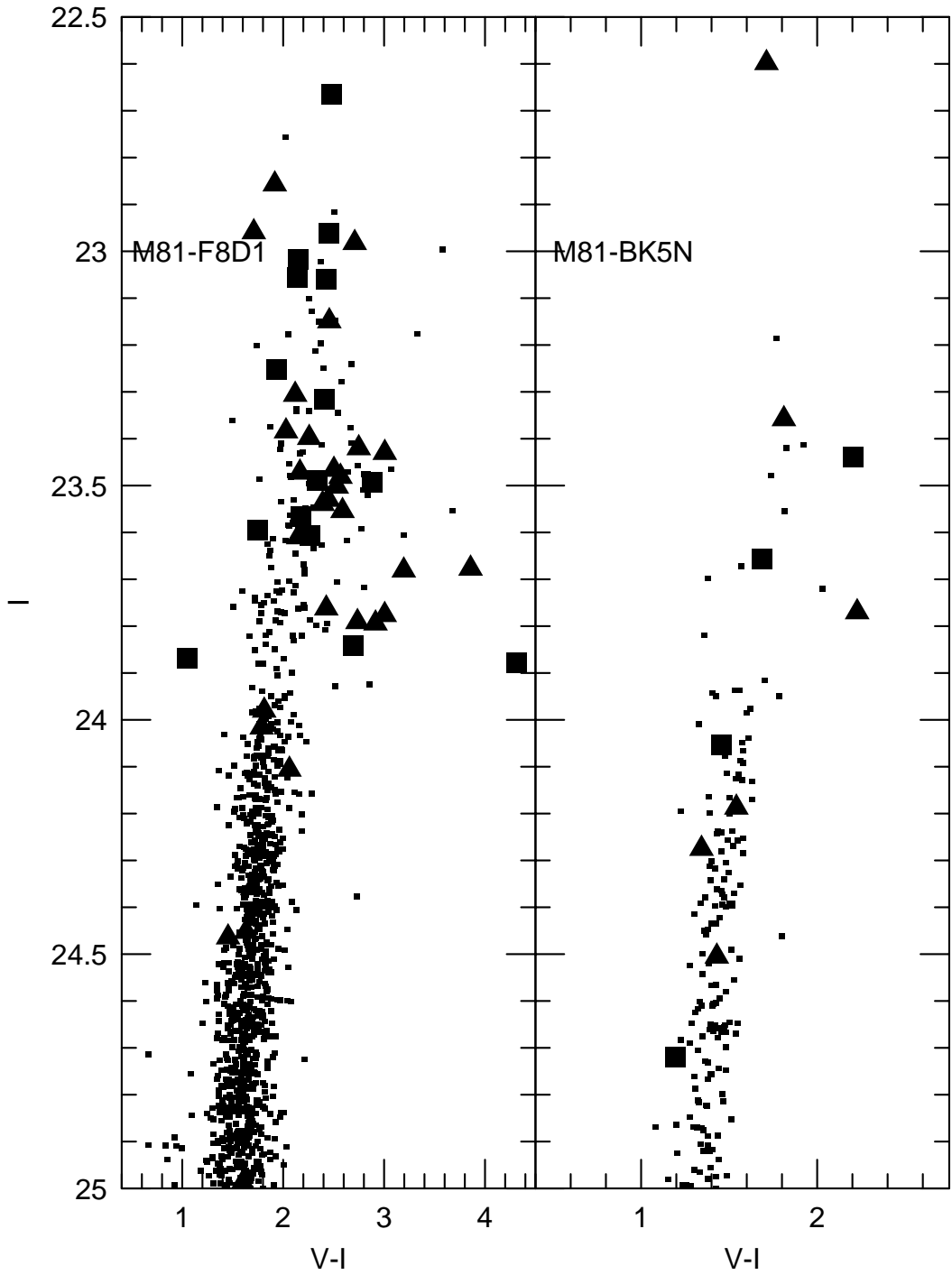}{7.0in}{0.}{80.}{80.}{-250.}{-48.}
\caption[]{Variables in BK5N and F8D1 identified in the c-m diagrams.
Certain variables are shown with filled squares, probable variables as
triangles.}
\label{variables_plot}
\end{figure}

We created a table of magnitude differences from the VI photometry for
each of the two epochs, and divided these differences by the formal
uncertainties in the differences.  These tables were searched for
magnitude differences greater than 4 standard deviations.  Stars were
flagged as certain variables if both V and I had changes of at least
4$\sigma$.  Stars with one color deviating by at least 4$\sigma$ but
the other less than 4$\sigma$ were flagged as probable variables.
Our criteria will likely not select all of the variables 
on our frames, but given the limitations of two epochs we feel it best
to be conservative.
Figure \ref{variables_plot} shows the c-m diagrams with the variables
identified.  The great majority of the variables are located, as
expected, near or above the red giant branch tip in both galaxies.  A
small number of variables, however, were found at luminosities
significantly fainter than the red giant branch tip (e.g.\ BK5N 1943-1
at I $\approx$ 24.7 and F8D1 1977-2 at I $\approx$ 25.6).  In total,
and here we include variables from the WF frames for BK5N and from the
PC frames for F8D1 which are not shown in Figs.\ \ref{bk5n_cmd} and
\ref{f8d1_cmd}, there are 4 certain and 8 probable variables in BK5N
and 18 certain and 38 probable variable stars in F8D1.  Table
\ref{variables_tab} lists the stars (the ID numbers are made up of a
four digit number to which is appended an integer indicating which CCD
chip the star is found in the first visit data), their photometry and
the magnitude differences and significances of variations.  Generally,
these stars become redder as they fade (which is normal; see for
example Eggen 1972, Lloyd-Evans 1983).

The variables that lie above the red giant branch tip are undoubtedly
members of an AGB population in both galaxies which will be discussed
in Sec.\ \ref{uagb}.  On the other hand, the variables with magnitudes
near the red giant branch tip are most probably analogous to the long
period variables found in globular clusters and the Galactic dE
companions (see, for example, Fig.\ 8 of Kaluzny et al.\ 1995).
Similarly, the fainter variables would probably be classified as yellow
semi-regular variables (SRd stars; Rosino 1978).

\section{Distance} \label{distance}

In recent years the technique of establishing distances to galaxies via
the determination of the luminosity of the red giant branch tip in the
I-band has become widely used (e.g.\ Gallart et al.\ 1996a, Minniti \&
Zijlstra 1996).  Provided the population under study has sufficient
metal-poor ([Fe/H] $\lesssim$ --0.7) red giants to allow an accurate
determination of the red giant branch tip I magnitude, this technique
has a precision similar to that for Cepheids (e.g.\ Lee et al.\ 1993a).
In the upper panel of Fig.\ \ref{lumin_func} we show the I-band
luminosity function (LF) for BK5N\@.  This function is based on the c-m
diagram of Fig.\ \ref{bk5n_cmd} and includes all stars fainter than I =
23.  No correction has been made for foreground Galactic stars since,
as Fig.\ \ref{field_cmd} shows, the number of non-member stars is
negligible.  This LF shows a distinct drop between the 23.9 $\le$ I
$\le$ 24.0 bin and the next brighter one, which we interpret as the
signature of the red giant branch tip.  We therefore adopt I(TRGB) =
23.95 $\pm$ 0.11 where the uncertainty given is the combination of the
uncertainty in the adopted value of I(TRGB), taken as $\pm$1 bin width
($\pm$0.10 mag), and the $\pm$0.05 mag zeropoint uncertainty assumed
for the photometry.  Using the reddening value given in Table
\ref{phot_table}, we then have I$_{0}$(TRGB) = 23.86 $\pm$ 0.12 where
now a $\pm$0.02 mag uncertainty in the adopted E(B--V) value is also
included.

Da Costa \& Armandroff (1990) give a relation between the absolute
bolometric magnitude of the red giant branch tip, based on the
horizontal branch luminosity scale of Lee et al.\ (1990), and mean
abundance.  They also give an equation for the bolometric correction to
the I magnitude of red giants as a function of (V--I)$_{0}$ color.
Using (V--I)$_{0}$ = 1.49 for the mean dereddened color at the giant
branch tip (cf.\ Fig.\ \ref{bk5n_cmd}), and using a mean abundance of
$<$[Fe/H]$>$ = --1.7 $\pm$ 0.25 (see next section), the Da~Costa \&
Armandroff (1990) relations yield \mbox{M$_{\rm I}$ = --4.0} and thus
(m--M)$_{0}$ = 27.9 $\pm$ 0.15 for BK5N\@.  Here the uncertainty now
also includes the effect of the uncertainty in the mean metal
abundance.

In the lower panel of Fig.\ \ref{lumin_func} the LF for F8D1 is shown.
This LF includes all stars fainter than I = 22.6 in
Fig.\ \ref{f8d1_cmd} except for a small number of outliers that lie
well to the blue of the red giant branch, though including these stars
would not alter the outcome.  Based on the field results for BK5N
depicted in Fig.\ \ref{field_cmd}, we assume that the contribution of
non-member stars to the F8D1 LF is also negligible.  It is immediately
evident from Fig.\ \ref{lumin_func} that I(TRGB) for F8D1 is $\sim$0.1 mag
fainter than it is for BK5N\@.  We adopt I(TRGB) = 24.0 $\pm$ 0.07
where, as for BK5N, the uncertainty is the combination of the
uncertainty in the location of I(TRGB), here $\pm$0.05 mag since the
larger number of stars permits a finer binning, and the assumed
$\pm$0.05 mag uncertainty in the photometric zeropoint.  Using the
reddening for F8D1 given in Table \ref{phot_table} and following the
same procedures as for BK5N, the distance modulus of F8D1 determined by
this method is (m--M)$_{0}$ = 28.0 $\pm$ 0.10.  The uncertainty given
includes the effects of uncertainty in the reddening and in the mean
metal abundance.  Of course neither this value for F8D1 nor that for
BK5N include any possible systematic uncertainty in the horizontal
branch distance scale on which the TRGB method is based.  Nevertheless,
these modulus values unequivocally associate both dwarfs with the M81
group, since Freedman et al.\ (1994) give (m--M)$_{0}$ = 27.8 $\pm$
0.20 for M81 itself based on Cepheid variables.  The apparent difference
in line-of-sight distance
between BK5N and F8D1 is, at $\sim$200 kpc, similar to their projected
separation on the sky of $\sim$140 kpc.  This (and Fig.\ \ref{pos_fig})
illustrates, if nothing else, the compactness of the M81 group compared
to the Local Group.

\section{Metal Abundances and Abundance Dispersions} \label{metallicity}

With the distance moduli of these dwarf galaxies established, we can
now derive their mean metal abundances and investigate any internal
abundance ranges via a comparison with the fiducial globular cluster
giant branches of Da~Costa \& Armandroff (1990).  These comparisons are
illustrated in Figs.\ \ref{bk5n_cmd} and \ref{f8d1_cmd}.  In their
paper Da~Costa \& Armandroff (1990) gave an abundance calibration based
on the mean (V--I)$_{0}$ color of the red giant branch at a luminosity
of M$_{\rm I}$ = --3 (which corresponds to I $\approx$ 25.0 for these
M81 group objects).  However, Armandroff et al.\ (1993) noted that the
sensitivity to abundance is increased, and the influence of AGB stars
decreased (such stars were specifically excluded in the Da~Costa \&
Armandroff (1990) calibration but cannot be here), if a luminosity of
M$_{\rm I}$ = --3.5 was adopted for the abundance determination (see
also Lee et al.\ 1993a).  The one half magnitude increase in the
luminosity employed also reduces the influence of photometric errors in
most situations.  Armandroff et al.\ (1993) presented a linear
relation, appropriate for [Fe/H] $\leq$ --1.3, between
(V--I)$_{0,-3.5}$ (the dereddened giant branch color at M$_{\rm I}$ =
--3.5) and [Fe/H] but illustrated only in their Fig.\ 7 the quadratic
relation appropriate for the full abundance range of the calibration
clusters \mbox{(--2.2 $\leq$ [Fe/H] $\leq$ --0.7)}.  That quadratic
relation, which we will employ here is:
\begin{center}
[Fe/H] = --1.00 + 1.97q -- 3.20q$^{2}$ 

\end{center}
where q = ((V--I)$_{0,-3.5}$ -- 1.6).  The rms deviation about the fit
is 0.08 dex.

Using the modulus derived in the previous section, M$_{\rm I}$ = --3.5
corresponds to I = 24.49 for BK5N.  Thus to determine
(V--I)$_{0,-3.5}$, the mean dereddened color for the stars in the
magnitude interval \mbox{24.39 $\le$ I $\le$ 24.59} was computed.
Excluding one star that lies $\sim$5$\sigma$ away from the mean and one
probable variable that lies in this magnitude interval, the remaining
25 stars yield \mbox{(V--I)$_{0,-3.5}$ = 1.360 $\pm$ 0.015} where the
listed uncertainty is the standard error of the mean.  Applying the
above calibration then yields the mean abundance of BK5N as
\mbox{$<$[Fe/H]$>$ = --1.7 $\pm$ 0.25}\footnote{As can be seen from
Fig.\ 7 of Armandroff et al.\ (1993), this value is essentially
identical with that which results from using the linear
((V--I)$_{0,-3.5}$, [Fe/H]) calibration.}.  Here the listed error
includes contributions from the calibration relation (0.08 dex), the
statistical error in (V--I)$_{0,-3.5}$ (0.05 dex), the uncertainty due
to zeropoint ($\pm$0.05 mag) and E(B--V) ($\pm$0.02 mag) uncertainties
(0.20 dex), and the effect of a $\pm$0.15 mag uncertainty in the
distance modulus (0.10 dex), with
the combined zeropoint and reddening uncertainties being dominant.
A consistency check on this derived
mean abundance is provided by applying the Da~Costa \& Armandroff
(1990) calibration at M$_{\rm I}$ = --3.0 (I = 24.99).  We find
an abundance higher by only 0.18 $\pm$ 0.14, where in this instance 
the uncertainty includes only the calibration and statistical uncertainties. 
This difference is acceptably small and indicates
that our adoption of $<$[Fe/H]$>$ = --1.7 $\pm$ 0.25 for BK5N is
unlikely to be significantly in error.

The standard deviation in (V--I) color for the sample of BK5N giants
used in the mean abundance determination is 0.075 mag, larger than that
expected from the errors in the photometry at this magnitude (0.050
mag).  Here it is important to note that the expected color error is
the same regardless of whether we consider the mean photon statistics
error for these stars, or the equivalent color error calculated on the
basis of comparing the two independent sets of photometry
(cf.\ Table \ref{error_tab}).  Subtracting (in quadrature) this error
contribution from the observed dispersion then yields an intrinsic
(V--I) color dispersion of \mbox{$\sigma_{\rm int}$(V--I) = 0.056 $\pm$
0.02}.  The error in this color dispersion is derived from its variation as
the sample changes due to shifts in the adopted modulus of order $\pm$0.15
mag.  Armandroff et al.\ (1993) noted that at this luminosity (M$_{\rm
I}$ = --3.5), the asymptotic and red giant branches in globular
clusters are essentially coincident, so for a single age population,
any intrinsic color dispersion can be interpreted as an intrinsic
abundance dispersion.  However, in BK5N there is evidence from the c-m
diagram that suggests the existence of a range of ages in this dwarf
(see following section).  As a result, the intrinsic color spread on
the giant branch will contain a component from the age distribution as
well as from the intrinsic abundance distribution.  Fortunately, the
effective temperature (and hence V--I color) of a red giant branch star
is much more sensitive to abundance than it is to the star's mass
(age).  Thus we are justified in interpreting the intrinsic color
dispersion primarily as the consequence of an intrinsic abundance
dispersion in BK5N, though the quantitative value should be recognized
as an upper limit on the true abundance dispersion.  With this caveat,
the abundance calibration then yields \mbox{$\sigma$([Fe/H]) = 0.20
$\pm$ 0.07} for BK5N.  This value of $\sigma$([Fe/H]) for BK5N is
consistent with those for Local Group dE's.  For example, Armandroff
et al.\ (1993) list 0.16 $\leq$ $\sigma$([Fe/H]) $\leq$ 0.24 for
Andromeda III and Da~Costa et al.\ (1996) give $\sigma$([Fe/H]) = 0.20
for And~I, dE's which bracket BK5N in luminosity.  Similarly, Suntzeff
(1993) lists $\sigma$([Fe/H]) values between 0.19 and 0.30 for the
Galactic dE's.  Thus again we have an indication that BK5N is very
similar in its properties to the Local Group dE galaxies.

For F8D1, M$_{\rm I}$ = --3.5 corresponds to I = 24.56 and so we have
used the interval \mbox{24.46 $\leq$ I $\leq$ 24.66} to determine
(V--I)$_{0,-3.5}$ for this galaxy.  Excluding one variable star,
we find that \mbox{(V--I)$_{0,-3.5}$ = 1.607} $\pm$
0.012.  The error is the standard error of the mean for the 183 stars
in the sample.  Applying the calibration given above then yields
$<$[Fe/H]$>$ = --1.0 $\pm$ 0.15 for F8D1.  As for BK5N, the uncertainty
given here includes contributions from the abundance calibration (0.08
dex), the statistical uncertainty from the sample (0.02 dex), the
effect of a $\pm$0.05 mag uncertainty in the zeropoint and a $\pm$0.02
mag uncertainty in E(B--V) (0.11 dex), and the result of altering the
distance modulus by $\pm$0.10 mag (0.08 dex).  Also as for BK5N, a
consistency check on this abundance is provided by calculating
(V--I)$_{0,-3.0}$ and applying the calibration of Da~Costa \&
Armandroff (1990).  We find an abundance higher by 
0.13 $\pm$ 0.11 dex, 
also with the error including only
the statistical and calibration uncertainties, 
a difference that is again satisfactorily small.
Moreover, it is apparent that the difference in mean abundance between
BK5N and F8D1, $\approx$0.7 dex, is independent of the luminosity on
the red giant branch at which the mean abundances are determined.

For the sample of stars used in determining (V--I)$_{0,-3.5}$ for F8D1,
the average photon statistics error in (V--I) is 0.083 mag while the
color error inferred from the repeat measurements is 0.086 mag.  Both
of these quantities are considerably less than the observed color
dispersion $\sigma_{\rm obs}$(V--I) = 0.160 $\pm$ 0.015.  Here the
uncertainty given reflects the change in $\sigma_{\rm obs}$(V--I) for
$\pm$0.10 mag changes in the distance modulus.  Adopting 0.085 mag as
the error contribution, we then have $\sigma_{\rm int}$(V--I) = 0.136
$\pm$ 0.02.  Once more interpreting this color dispersion as primarily
a reflection of an internal abundance range, the abundance calibration
yields $\sigma$([Fe/H]) = 0.26 $\pm$ 0.03 for F8D1.  Again as for BK5N,
given the likely existence of an age range in F8D1, this value is
strictly an upper limit on the true abundance dispersion.

It is illustrative to compare this value for the abundance dispersion
in F8D1 with values for Local Group dwarf systems.  For example, Sagar
et al.\ (1990) give $\sigma$([Fe/H]) = 0.31 $\pm$ 0.05 for the Fornax
dE from the intrinsic (B--V) width of the giant branch under the
assumption of no internal age variation.  Alternatively, they give
$\sigma$([Fe/H]) = 0.27 $\pm$ 0.06 if there is an age variation of
between 3 and 17 Gyr in this dE.  These values are similar to that
found for F8D1 which is somewhat more luminous, though of lower surface
brightness, than Fornax.  The other Local Group object with which
comparison seems appropriate is the dE galaxy NGC~147, which is
$\sim$0.7 mag more luminous than F8D1.  Han et al.\ (1997) have studied
NGC~147 in detail using HST observations and have employed a similar
analysis to that used here; a direct comparison is then possible.  At
\mbox{M$_{\rm I}$ = --3.0}, Han et al.\ (1997) give $\sigma_{\rm
int}$(V--I) = 0.13 for their outer NGC~147 field (the same value as
found by Mould et al.\ 1983 from ground-based data at a larger radial
distance), while for F8D1 we find $\sigma_{\rm int}$(V--I) = 0.11 $\pm$
0.02 (at M$_{\rm I}$ = --3.0).  Using the Da~Costa \& Armandroff (1990)
calibration at the appropriate (V--I)$_{0,-3.0}$ color, these values
translate to $\sigma$([Fe/H]) = 0.42 and $\sigma$([Fe/H]) = 0.27 $\pm$
0.04 for NGC~147 and F8D1, respectively.  It appears then that the
abundance distribution in F8D1 is somewhat narrower than it is in the
outer parts of NGC~147.  Of course it has to be kept in mind that the
presence of an age range in both galaxies (Han et al.\ 1997; section 7)
means that these abundance dispersions are strictly upper limits.  It
is, however, reassuring to note that the value of $\sigma$([Fe/H])
derived for F8D1 at M$_{\rm I}$ = --3.0 agrees with that found at the
higher luminosity.

One final point deserves comment here.  Although the F8D1 data shown in
the c-m diagram of Fig.\ \ref{f8d1_cmd} do not cover a substantial
fraction of the galaxy's extent, it is nevertheless of interest to
investigate whether there is any indication of a radial gradient in the
mean giant branch color.  To do this we have split the F8D1 sample into
an inner group (r $<$ 500 pix from the center, corresponding to
50$\arcsec$ which is 0.6 core radii or 0.4 R$_{\rm eff}$) and an outer
group (r $>$ 750 pix, equivalent to 75$\arcsec$, 0.9 core radii or 0.6
R$_{\rm eff}$).  For the inner group the mean dereddened color of the
giant branch at M$_{\rm I}$ = --3.5 is 1.585 $\pm$ 0.025 (49 stars),
while for the outer group (V-I)$_{0,-3.5}$ = 1.575 $\pm$ 0.020 (51
stars).  The difference (inner$-$outer) of 0.010 $\pm$ 0.032 mag thus
offers no support for the existence of a radial abundance gradient in
F8D1, at least within $\sim$1 core radius.  The 3$\sigma$ upper limit
on any possible change in abundance between the two regions is 0.20
dex.  Similarly, the BK5N sample has been split into two groups
divided at a (geometric) radius of 330 PC pixels, or $\sim$15$\arcsec$
which is approximately 1 core radius or 0.8 R$_{\rm eff}$.  For the
inner sample (V-I)$_{0,-3.5}$ = 1.353 $\pm$ 0.017 (20 stars) while for
the outer sample (V-I)$_{0,-3.5}$ = 1.383 $\pm$ 0.031 (6 stars).  As
for F8D1, the difference in mean giant branch color, -0.030 $\pm$ 0.035
mag, does not support the existence of any radial abundance gradient in
BK5N\@.  The corresponding 3$\sigma$ upper limit on any possible change
in abundance between these two groups is $\sim$0.4 dex.  These results
are consistent with those for the Local Group galaxies NGC~147 (Han et
al.\ 1997) and And~I (Da~Costa et al.\ 1996) where in neither case was
any strong evidence found to support the existence of a radial
abundance gradient.

\section{Upper Asymptotic Giant Branch Stars} \label{uagb}

It is immediately apparent from Figs.\ \ref{bk5n_cmd} and
\ref{f8d1_cmd} that both these galaxies contain significant numbers of
stars that lie above the red giant branch tip.  The lack of stars
brighter than I $\approx$ 24 in the ``field'' regions of
Fig.\ \ref{field_cmd} strongly suggests that the majority, if not all,
of these stars are members of their respective galaxies.  The high
incidence of variability among this population
(cf.\ Fig.\ \ref{variables_plot}) is a further testament to the membership
of these stars.  Indeed, in BK5N 5 out of 14 (36\%) stars brighter than I =
23.8 are listed as probable or certain variables while 39 out of 158
(25\%) stars brighter than I = 23.9 in F8D1 are similarly listed.
Undoubtedly these variable star fractions are lower limits on their
true values, given that we have data at only two epochs.

The existence of stars with luminosities above that of the red giant
branch tip, i.e.\ upper-AGB stars, is, in populations with [Fe/H]
$\lesssim$ --1.0, an unambiguous signature of the presence of an
intermediate-age population, a
population whose age is considerably less than that of the Galactic
globulars.  Generally, on an age scale where the Galactic globular
clusters are $\sim$13 -- 15 Gyr old, an intermediate-age population is
taken to be one whose age is less than $\sim$10 Gyr.  The luminosity
that an AGB star reaches before it sheds its remaining envelope as a
planetary nebula, with the core becoming a white dwarf, is dependent on
the envelope mass.  All other things being equal, a younger star will
have a larger envelope mass and can then evolve to higher luminosities
on the AGB than an older star.  Thus the maximum luminosity observed on
the upper-AGB can serve as an age indicator.  There are, however, two
things that must be kept in mind when using this technique.  First,
since the AGB termination luminosity is a function of envelope mass,
anything that changes the envelope mass will affect the luminosity
evolution.  In particular, these low surface gravity, high luminosity
stars will undergo significant mass loss on the AGB.  Theoretically,
this mass loss can be parameterized in a number of ways, but the details
of the evolution will be sensitive to the adopted mass loss law (see,
for example, Appendix A of Gallart et al.\ 1996b).  Second, upper-AGB
evolution is rapid so that a reasonable sample of upper-AGB stars is
required in order to provide a valid estimate of the AGB termination
luminosity.

We will be concerned with estimating two quantities for these dwarf
galaxies from our data: first, a limit on the age of the
intermediate-age population from the luminosity of the brightest
upper-AGB stars, and second, an estimate of the fraction of the total
population of each dwarf that exists as intermediate-age stars.  As
discussed, for example, by Da~Costa (1997), the Local Group dE
galaxies show a large and surprising variety of intermediate-age
population fractions, from essentially none in systems such as Ursa
Minor to substantial in systems such as Carina, Fornax and Leo I\@.  We
begin with BK5N.

\subsection{BK5N} \label{uagb_bk5n}

For BK5N, the mean metal abundance and metal abundance dispersion
derived in Sec.\ \ref{metallicity} suggest that it is unlikely that
there are any stars in this dwarf galaxy with abundances exceeding
[Fe/H] $\approx$ --1.0 (see also Fig.\ \ref{bk5n_cmd}).  Consequently,
if the stellar population of BK5N consisted entirely of stars
comparable in age to the Galactic globular clusters, then we would not
expect to see any stars with luminosities above that of the red giant
branch tip (e.g.\ Frogel \& Elias 1988, Renzini 1992).  Clearly this is
not the case in Fig.\ \ref{bk5n_cmd} and we can immediately conclude
that BK5N {\it contains an intermediate-age population}.

Using the bolometric corrections to I-band magnitudes from Da~Costa \&
Armandroff (1990) and the distance modulus from Sec.\ \ref{distance},
we have calculated bolometric magnitudes for the 14 stars brighter than
I = 23.8 in Fig.\ \ref{bk5n_cmd}.  With the exception of one star at
M$_{\rm bol}$ $\approx$ --4.9 which we will discuss shortly, the
luminosity function of the other stars indicates that the AGB
termination luminosity in BK5N is M$_{\rm bol}$ $\approx$ --4.3; 5 of
the 13 stars have bolometric magnitudes at, or $\leq$0.2 mag fainter
than, this value.  There are two approaches to interpreting this value:
empirical and theoretical.  Empirically, we note that the value for
BK5N is fainter than the luminosity of the brightest (upper-AGB) carbon
stars in the Leo~I and Carina Local Group dE's, for which Azzopardi
(1994) gives M$_{\rm bol}$ $\approx$ --4.5 and M$_{\rm bol}$ $\approx$
--4.6, respectively.  Both these dE galaxies have substantial
intermediate-age populations containing stars as young as a few Gyr
(e.g.\ Lee et al.\ 1993c, Smecker-Hane et al.\ 1994).  On the other hand
Azzopardi (1994) lists M$_{\rm bol}$ $\approx$ --4.4 for the brightest
carbon stars in Leo II\@.  Mighell \& Rich (1996) have concluded that
this dE galaxy contains only stars older than $\sim$7 Gyr with a
``typical'' Leo II member having an age of 9 Gyr.  Similarly, the
brightest carbon star in the SMC cluster Kron 3, which has an age of
$\sim$9 Gyr (Rich et al.\ 1984, Alcaino et al.\ 1996), has a luminosity
of M$_{\rm bol}$ $\approx$ --4.2 (Frogel et al.\ 1990).  All these
stellar systems have mean abundances that are similar to BK5N\@.  Thus
it seems reasonable to conclude that BK5N, like many of the Local Group
dE's, has had an extended period of star formation, forming stars
until $\sim$8 Gyr ago.

The single bright star (at I $\approx$ 22.6, V--I $\approx$ 1.7) in the
BK5N c-m diagram, if a member, offers a somewhat different
interpretation.  This star is well isolated in the c-m diagram which
might suggest non-membership.  However, it is also a probable variable
in that the V magnitudes for the two epochs differ by 0.17 mag, though
the I magnitudes differ by only 0.01 mag.  If it is a member, then it
has M$_{\rm bol}$ $\approx$ --4.9, which, interpreted as an AGB
termination luminosity, suggests an age of 2 -- 3 Gyr.  Unfortunately,
there is no way at the present time we can definitely establish the
membership of this star in BK5N\@.

The context for a theoretical interpretation of these results is
provided by the Bertelli et al.\ (1994) isochrones, since these include
the evolution of stars through the upper-AGB phase.  However, it is not
clear that at the metallicities appropriate for BK5N these isochrones
offer much insight.  As noted above, the AGB termination luminosity is
dependent on the envelope mass, which is in turn determined by the mass
loss rate.  In the Bertelli et al.\ (1994) isochrones the mass loss
rate at a fixed luminosity is a strong function of the effective
temperature (i.e.\ radius) of the AGB star.  Thus the relatively blue
metal-poor AGB stars, as parameterized, undergo less significant mass
loss as they ascend the AGB\@.  Consequently, and particularly for the
most metal-poor case which corresponds to the mean abundance of BK5N,
the theoretical AGB tip luminosities are at M$_{\rm bol}$ $\approx$
--4.8 even for ages of 8 to 10 Gyr.  Indeed, even for globular cluster
like ages, the metal-poor Bertelli et al.\ (1994) isochrones have AGBs
that extend well above the red giant branch tip, in conflict with the
observations.  Thus we have not used these isochrones to draw any
inferences, from a theoretical viewpoint, concerning the BK5N upper-AGB
population.  It is worth noting though that the Marigo et al.\ (1996)
calculations do suggest, from their Fig.\ 2, an age in excess of 6 Gyr
for the BK5N upper AGB stars, since the observed AGB termination
luminosity is fainter than M$_{\rm bol}$ $\approx$ --4.5.  The
calculations, however, were carried out at log(Z/Z$_{\sun}$) = --0.4,
considerably higher than the mean abundance of BK5N\@.

As for estimating the relative contribution of the intermediate-age
population to BK5N's total luminosity, we follow the procedures
outlined in Armandroff et al.\ (1993).  This procedure makes use of the
formalism developed by Renzini \& Buzzoni (1986) for the V band in
which the number of AGB stars expected from a total population
characterized by a total luminosity
 L$_{\rm V}$ is given by:
\begin{center}
N$_{\rm AGB}$ = 4 $\times$ 10$^{-5}$ L$_{\rm V}$ t$_{\rm AGB}$
\end{center}
Here N$_{\rm AGB}$ is the number of upper-AGB stars per magnitude and
t$_{\rm AGB}$ is the upper-AGB evolutionary rate in Myr/mag.  Using
t$_{\rm AGB}$ = 1.3 (Mould 1992) and M$_{\rm V}$ = --11.3 (reduced by
the fraction of the BK5N total luminosity contained on the PC frames,
which is one-third), the 13 upper-AGB stars between 23.8 $\geq$ I
$\geq$ 22.8 yield an intermediate-age population fraction for BK5N of
0.3 $\pm$ 0.1.  The error is simply the statistical error in the sample
of BK5N upper-AGB stars.  This intermediate-age population fraction is
considerably larger than the 0.1 $\pm$ 0.1 fractions calculated using
this technique for the M31 dE companions And~I and And~III by
Armandroff et al.\ (1993).  It suggests that BK5N is perhaps analogous
to other Local Group systems such as Carina and Leo~I which have
substantial intermediate-age populations.  For example, in Carina the
intermediate-age population fraction is directly observed to be of
order 0.7 to 0.8 (Mighell 1990, Smecker-Hane et al.\ 1994).  Clearly,
whatever the mechanism for producing the significant extended star
formation histories seen in these galaxies is, it is not restricted to
Local Group dwarfs.

\subsection{F8D1} \label{uagb_f8d1}

\begin{figure}[p]
\plotone{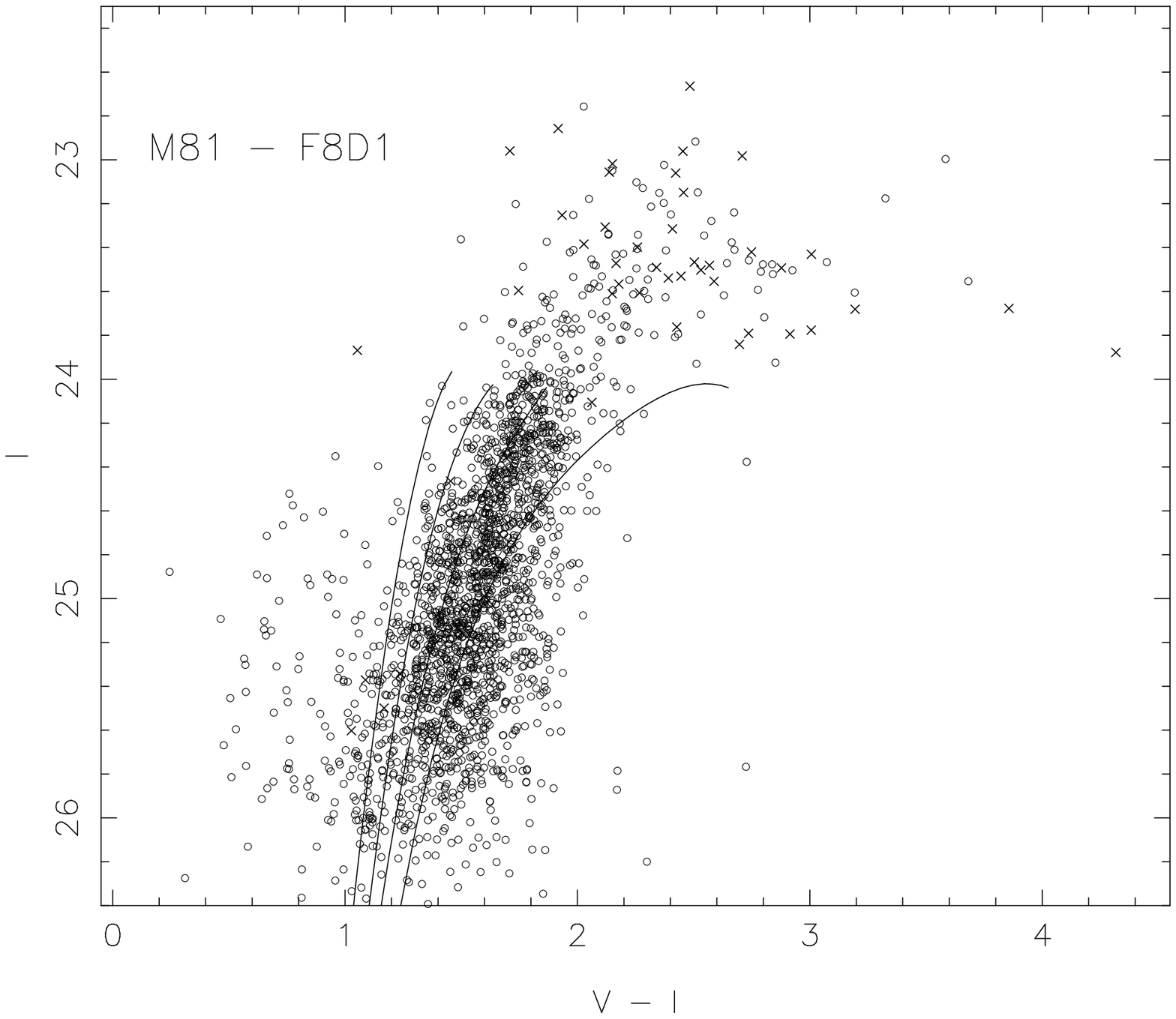}
\caption[]{Color--magnitude diagram for M81 F8D1 (WF frames only).
Standard globular cluster giant branches are shown for M15 ([Fe/H] =
--2.17), M2 (--1.58), NGC 1851 \mbox{(--1.16),} and 47 Tuc
(--0.71) (Da Costa \& Armandroff 1990) shifted to the distance modulus
and reddening of F8D1 \mbox{(($m$--$M$)$_0$} = 28.0; E($V$--$I$) =
0.043).  The $\times$ symbols represent stars with significant
photometric variations between the two visits.}
\label{f8d1_cmd}
\end{figure}

\begin{figure}[p]
\plotone{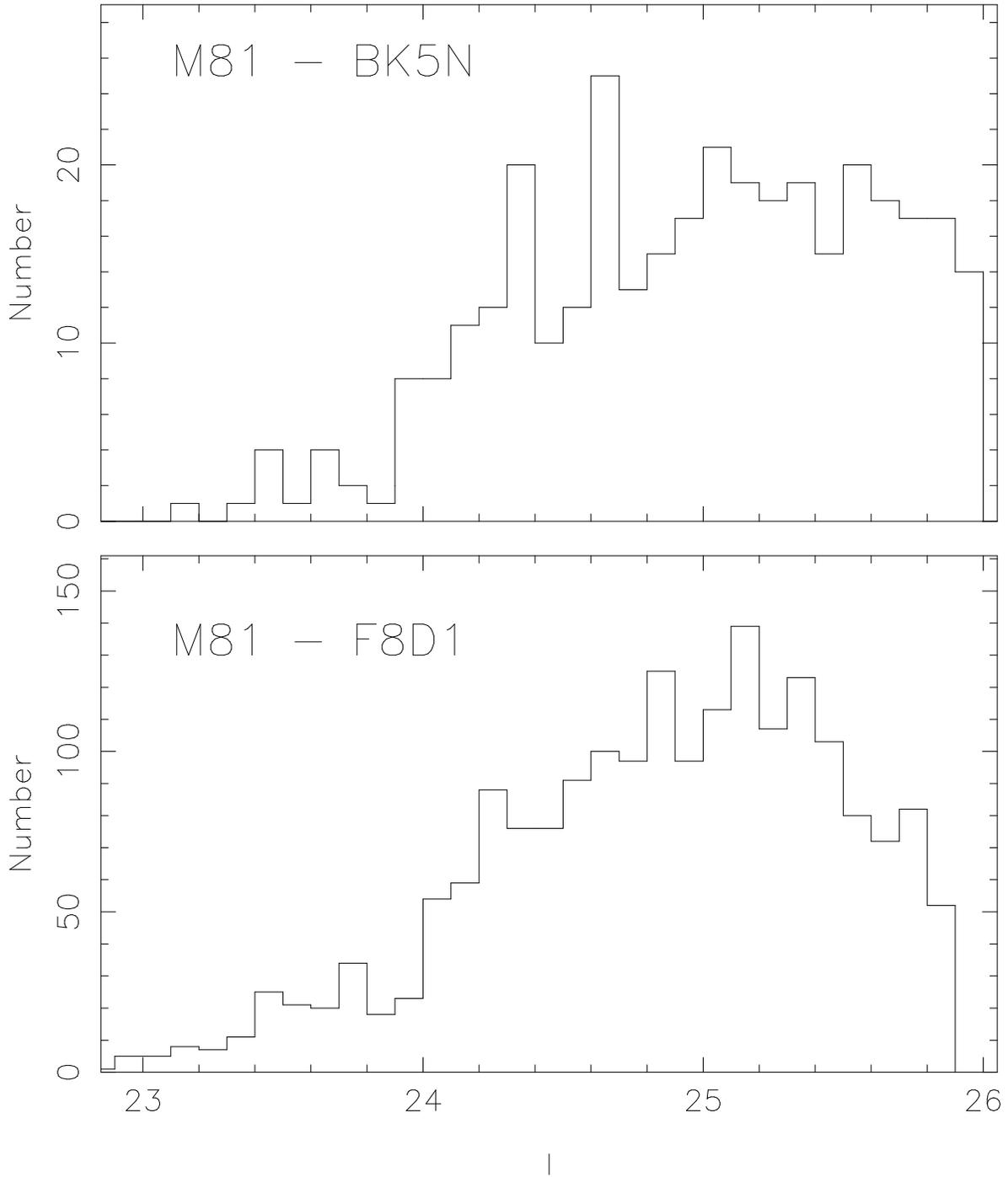}
\caption[]{I-band luminosity functions for BK5N (upper) and F8D1
(lower) using 0.1~mag bin sizes.  These luminosity functions are based
on the c-m diagrams shown in Figs.\ \ref{bk5n_cmd} and \ref{f8d1_cmd}
and have a negligible contribution from foreground stars. The red giant
branch tips are evident as distinct drops in the LFs at I $\approx$
23.9 (BK5N) and I $\approx$ 24.0 (F8D1).  Note that the F8D1 red giant
branch tip magnitude was actually determined from a luminosity function
that used a 0.05 mag bin size.}
\label{lumin_func}
\end{figure}

As for BK5N we have used the I-band bolometric corrections from
Da~Costa \& Armandroff (1990) and the distance modulus from
Sec.\ \ref{distance} to calculate bolometric magnitudes for the bright red
stars in Fig.\ \ref{f8d1_cmd}.  The most luminous of these stars have
M$_{\rm bol}$ $\approx$ --5.0 (7 stars within $\pm$0.2 mag of this
value) while the bolometric luminosity function shows a substantial
rise at M$_{\rm bol}$ $\approx$ --4.4 (38 stars with --4.4 $\le$
M$_{\rm bol}$ $\le$ --4.2 versus 34 stars with --5.2 $\le$ M$_{\rm
bol}$ $\le$ --4.4).  These bolometric magnitudes are again similar to
those for the more luminous stars in Local Group dE galaxies.  Among
the Galactic companions Fornax has upper-AGB carbon stars as luminous
as M$_{\rm bol}$ $\approx$ --5.6 (Azzopardi 1994) while the M31
companion And~II has upper-AGB stars with M$_{\rm bol}$ $\approx$ --4.5
(Aaronson et al.\ 1985).  The brighter upper-AGB carbon stars
identified by Richer et al.\ (1984) in NGC~205 also have M$_{\rm bol}$
$\approx$ --5.0.  However, given the F8D1 mean metal abundance and
abundance dispersion (cf.\ Sec.\ \ref{metallicity}), interpreting the
luminosities of these F8D1 stars in terms of age is not straightforward
--- we cannot conclude, as was done for BK5N, that all the F8D1 stars
above the red giant branch tip have [Fe/H] $\leq$ --1.0 and that
therefore they must be of intermediate-age (cf.\ Frogel \& Elias 1988,
Guarnieri et al.\ 1997).

It is worth illustrating this point by considering the Galactic
globular cluster 47 Tuc ([Fe/H] = -0.7).  This cluster contains at
least 4 long period variable (LPV) AGB stars whose luminosities at
maximum reach as bright as M$_{\rm bol}$ $\approx$ --4.8 (Frogel et
al.\ 1981, adjusted to the distance modulus used in Da~Costa \&
Armandroff 1990), well above the red giant branch tip.  Further, even
at minimum light, they remain somewhat above the red giant branch tip
(Frogel et al.\ 1981).  Yet there is no reason to suggest that these
stars are any different in age from the rest of the cluster
population.  To further illustrate this point, 47 Tuc V2 near maximum
light would have I $\approx$ 22.7, V--I $\approx$ 1.90, if it were at
the distance of F8D1 (Lloyd-Evans 1983). These values are comparable to
the brightest F8D1 stars in Fig.\ \ref{f8d1_cmd}.

As a result, in order to investigate the possibility that F8D1 has an
intermediate-age population, we must first estimate the possible
numbers of luminous metal-rich, but old stars.  Table \ref{phot_table}
gives the total visual luminosity of F8D1 as M$_{\rm V}$ = --14.25,
though only $\approx$20\% of this luminosity is contained on the
overlapping WF frames from which Fig.\ \ref{f8d1_cmd} is drawn.   Given
the mean abundance and abundance dispersion from
Sec.\ \ref{metallicity}, we assume that one third of the luminosity of
the galaxy comes from stars with [Fe/H] $\gtrsim$ --0.9.  Such a
population has the capacity to produce LPVs above the red giant branch
tip even if it consists entirely of old stars.  We assume that such an
old population will produce stars above the red giant branch tip at the
same rate as 47 Tuc.  This cluster apparently contains 5 AGB LPV stars
(V1, V2, V3, V4 and V8; Frogel et al.\ 1981, Da~Costa \& Armandroff
1990) and the total visual luminosity of the cluster is M$_{\rm V}$
$\approx$ --9.5.  Together these assumptions predict $\sim$26 stars
above the red giant branch tip in the fraction of F8D1 represented by
Fig.\ \ref{f8d1_cmd}.  As noted above, these stars could have single
epoch luminosities anywhere between M$_{\rm bol}$ =  --3.8 and --4.8.
However, in Fig.\ \ref{f8d1_cmd} there are 147 stars brighter than
M$_{\rm bol}$ = --3.8.  Thus, despite the concerns regarding luminous
stars generated by an old metal-rich population, it does appear that
there is {\it a significant intermediate-age population in F8D1}, just
as there is in BK5N and in many of the Local Group dE galaxies.

With M$_{\rm bol}$ $\approx$ --5.0, the most luminous stars in F8D1
probably have ages of a few Gyr using either the LMC cluster data
(e.g.\ Frogel et al.\ 1990, Mould \& Da Costa 1988) for calibration, or
through noting that the Galactic companion dE's Fornax, Carina and Leo~I
have carbon stars of comparable luminosity together with populations of
this age, identified from main sequence turnoff magnitudes (see, for
example, Da~Costa 1997).  On the other hand, the brightest stars in
F8D1 are $\sim$0.7 mag brighter than those in BK5N (putting aside the
one bright BK5N star mentioned previously), and thus it would appear
that the epoch of last significant star formation is different by
perhaps 4 -- 5 Gyr in these two M81 group dE galaxies.  Again this
highlights that the diversity of star formation histories seen in Local
Group dE's is not purely a ``local'' phenomenon, it clearly also extends
to these M81 group dE's.

\begin{figure}[p]
\plotone{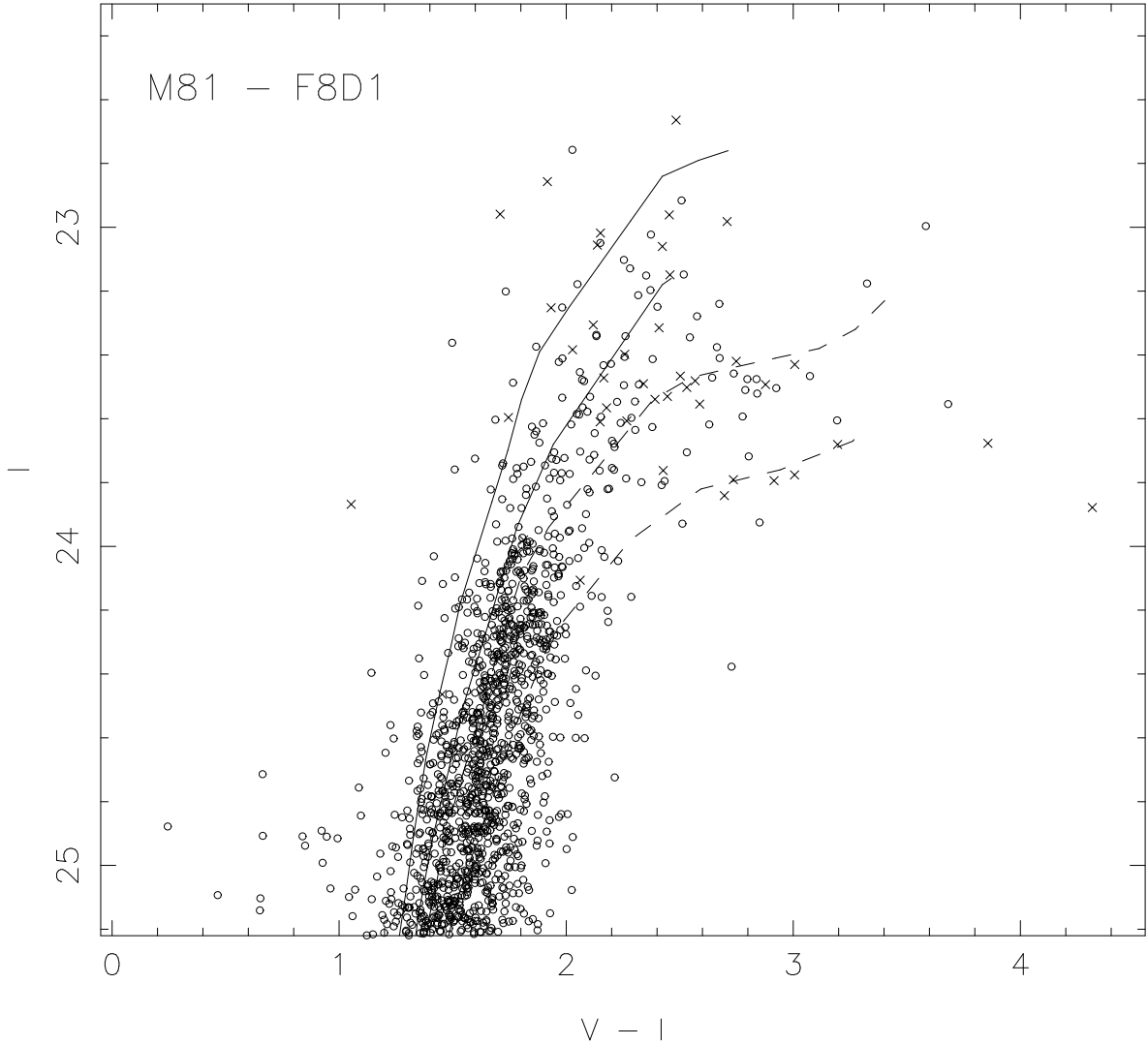}
\caption[]{Color--magnitude diagram for M81 F8D1.  Isochrones from
Bertelli et al.\ (1994) including AGB stars for log($Z$/$Z_{\odot}$) of
--1.3 (solid lines) and --0.7 (dashed lines) and ages of 4 and 10 Gyr
are superposed. At each abundance the younger isochrones reach higher
luminosities.  The $\times$ symbols represent stars with significant
photometric variations between the two visits.}
\label{f8d1_cmd_agb}
\end{figure}

We show in Fig.\ \ref{f8d1_cmd_agb} the F8D1 observations superposed
with AGB isochrones from the Bertelli et al.\ (1994) set.  Evidently,
at the relatively higher metallicity of F8D1, these isochrones do not
suffer to the same extent
from the problems described above in the case of BK5N.  Two
metallicities, which bracket the F8D1 mean abundance, are depicted as
are two ages for each abundance.  These data reinforce, as do the
calculations of Marigo et al.\ (1996), our conclusion that F8D1
contains stars with ages as young as 3 -- 4 Gyr.  It is equally
apparent though that a significant range of (intermediate) ages is
present.  The isochrones also show that age and abundance are to a
large degree degenerate for the majority of the F8D1 stars above the
red giant branch tip.  Observations in the near-IR could help break
this degeneracy and thus provide more insight into the extended star
formation history of this dE galaxy.

As regards an estimate of the intermediate-age population fraction in
F8D1, we again apply the equation given in Sec.\ \ref{uagb_bk5n}.  The
bolometric luminosity function for F8D1 has 102 stars/mag over the
interval --5.0 $\le$ M$_{\rm bol}$ $\le$ --4.0 which, after correcting
for the possible old metal-rich contribution and the fraction of the
total visual luminosity represented in Fig.\ \ref{f8d1_cmd}, yields an
intermediate-age fraction of 0.2.  Alternatively, as noted above, there
is a ``step'' in the F8D1 upper-AGB luminosity function at M$_{\rm bol}$
$\approx$ --4.4.  If we restrict the number of stars to those between
--4.4 $\le$ M$_{\rm bol}$ $\le$ --4.0, then after the same corrections,
the intermediate-age fraction estimate for F8D1 is 0.35.  The
difference between these two values is undoubtedly a measure of the
uncertainty in this technique.  However, it is clear that the
intermediate-age fraction in F8D1 is higher than that for the Local
Group dE's And~I and And~III, calculated with this technique (Armandroff
et al.\ 1993).  Further, there does not appear to be any significant
difference between F8D1 and BK5N in this respect, though as noted
above, the age of the youngest intermediate-age stars in F8D1 appears
to be approximately one-half that in BK5N\@.

\section{Globular Clusters} \label{gc}

Globular clusters in dwarf ellipticals have received more attention in
recent years than previously.  For example, Durrell et al.\ (1996a,b)
have studied globular clusters in the dwarf companion to NGC 3115 as
well as in a number of dE galaxies in the Virgo cluster. They concluded
that the mean specific frequency of occurrence of clusters in the dE's
is between 3 and 8, similar to that of giant ellipticals and higher
than that of spirals.  In accord with this result, the M31 companions
NGC~147, NGC~185 and NGC~205 have typically a half dozen or so
confirmed globular clusters associated with them (e.g.\ Da~Costa \&
Mould 1988).  Among the less luminous Local Group dE's, the Fornax
dwarf has 5 clusters and Sagittarius has 4, but none of the fainter
dE's have any known clusters.  Assuming then that dE's have 4 clusters
per M$_{\rm V}=-15$ in luminosity, we would expect to find around 2
globulars in F8D1, but none in BK5N.  Galactic globulars have
half-light radii in the range of 2 to 10 pc (see e.g., Djorgovski \&
Meylan 1994), so we would expect any globulars in our two dwarfs to
have half-light radii in the range of 0.1 to 0.5\arcsc.  Thus such
objects would be difficult to distinguish from stars in ground-based
data, but such a task lies within the capabilities of HST.

We searched for globulars in these two galaxies by first checking the
HST frames for appropriate resolved objects, and then measuring their
colors. Cluster candidates were those objects whose colors and
morphologies matched what we expect for clusters at the distance of
these two galaxies.  For F8D1 we are limited by the area covered by the
HST images: only about 3/4 of the area out to R$_{\rm eff}$ was
observed (or about 15\% of the area out to the limiting radius of the
light profile) --- an important shortcoming since the Fornax dwarf, for
example, has some globulars out to 2R$_{\rm eff}$.  Being much smaller,
BK5N's areal coverage was about 80\% out to the limiting radius.

\begin{figure}[p]
\plotone{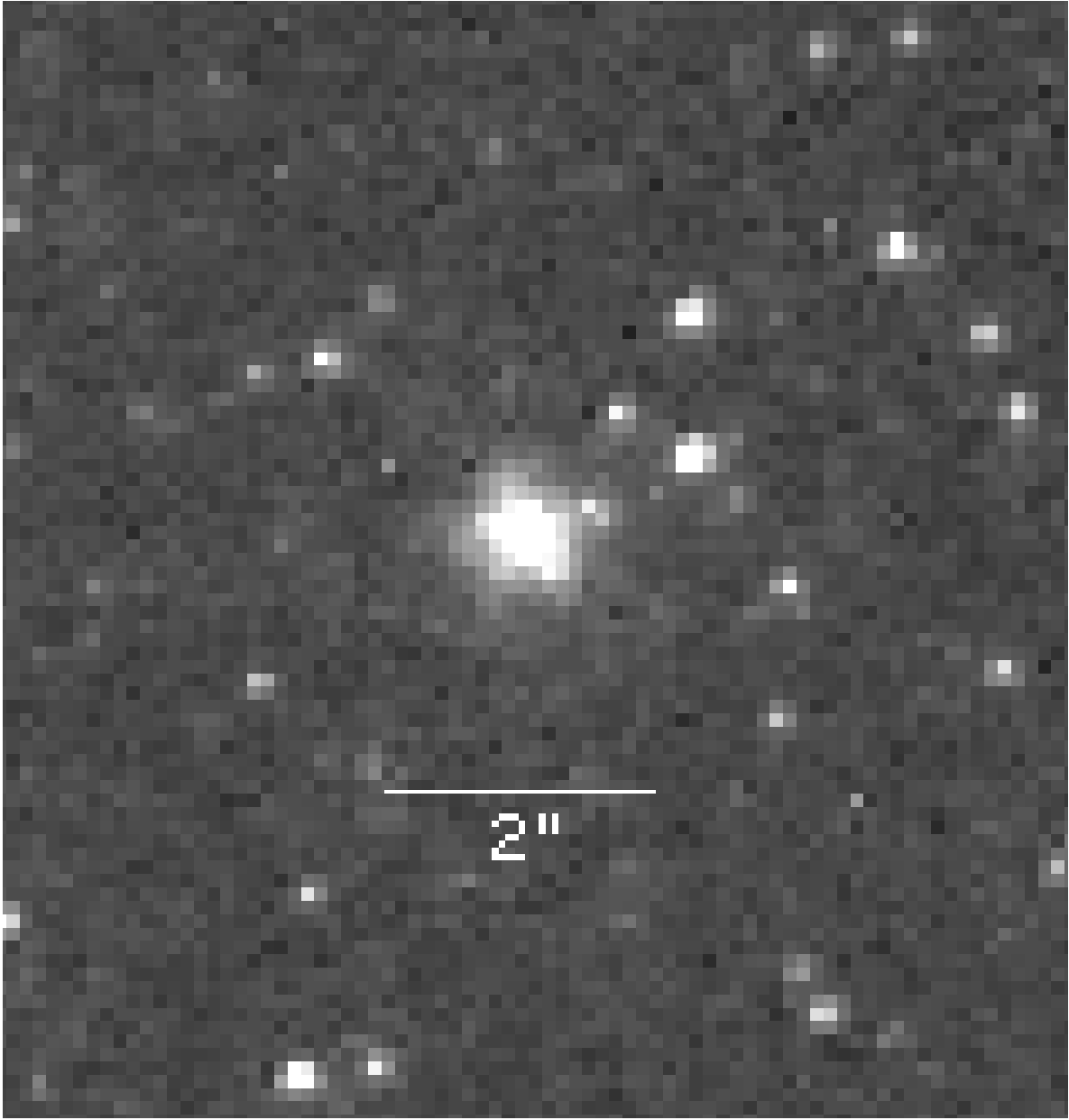}
\caption[]{Candidate globular cluster in F8D1, shown in the 
F814W filter image from HST.}
\label{glob_pic}
\end{figure}

While no likely candidates were found in the images of BK5N, several
candidates were identified in F8D1, one of which has both the color
expected for a globular, and an appearance that matches that of a star
cluster more than that of a background galaxy. In fact, it appears that
stars in this object, which we have designated F8D1-GC1, are partially
resolved. An enlargement of the HST image of this candidate is
presented in Fig.\ \ref{glob_pic}, and Table \ref{glob_tab} lists its
basic data.  The absolute magnitude and dereddened color ((V--I)$_0$
$\approx$ 0.72) of F8D1-GC1 are quite similar to those for the
metal-poor ([Fe/H] $\approx$ --1.80) Galactic globular cluster
NGC~4147, for example, for which the tabulation of Peterson (1993)
yields M$_{\rm V}$ $\approx$ --6.1 and (V--I)$_0$ $\approx$ 0.76.  If
we interpret the relatively blue (V--I)$_0$ integrated color of
F8D1-GC1 as indicating a low metal abundance (the alternative of a
younger age seems less likely, though additional integrated colors are
required to verify our assumption), then, given the mean metal
abundance of the F8D1 field stars derived in Sec.\ \ref{metallicity},
we have a further example of the result that the abundances of globular
clusters in dE galaxies are generally significantly lower than that of
the constituent field stars (Da~Costa \& Mould 1988, Durrell et
al.\ 1996a,b).

\begin{figure}[p]
\plotfiddle{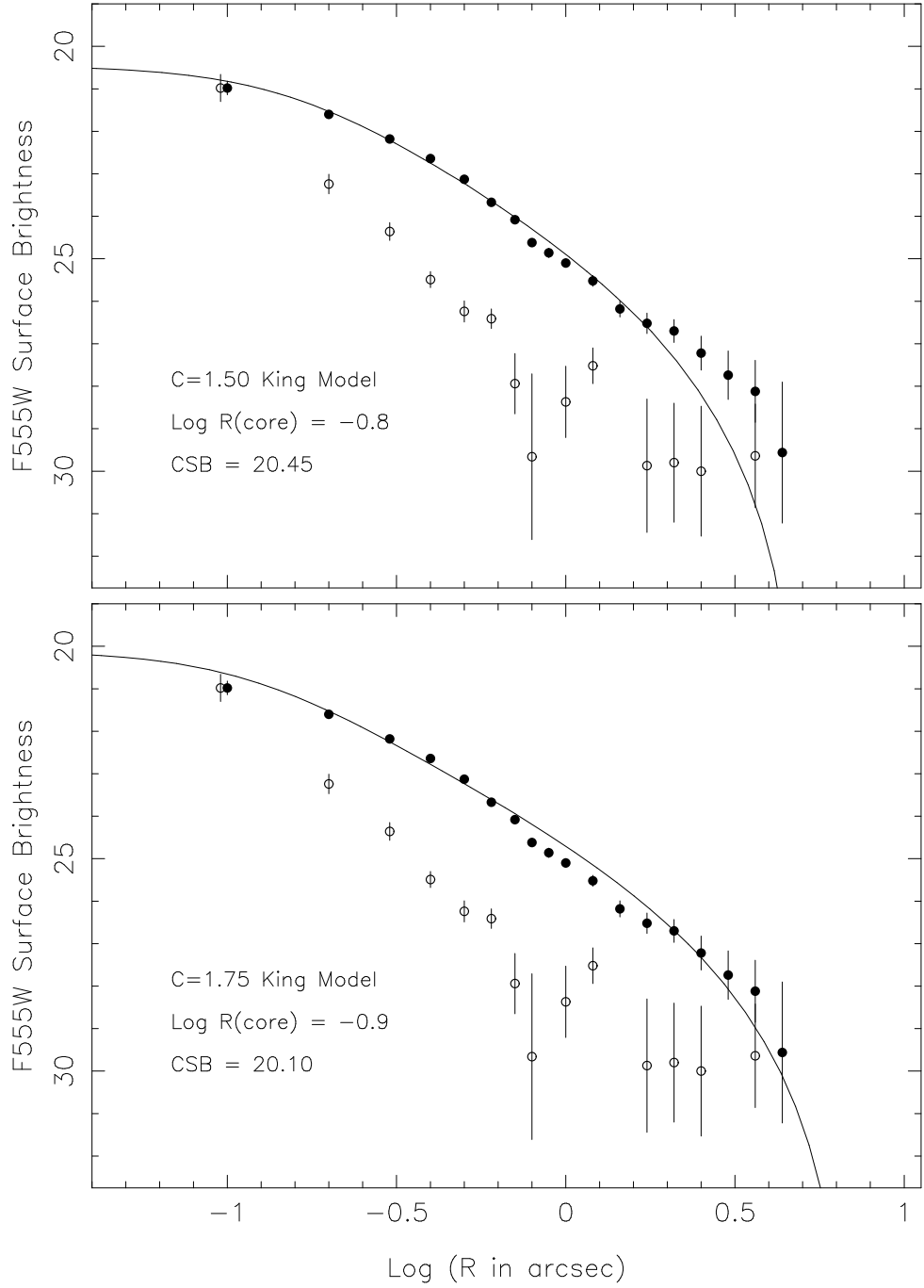}{6.9in}{0.}{80.}{80.}{-237.}{-66.}
\caption[]{Radial V light profile of the candidate globular cluster in
F8D1 (solid symbols).  The open symbols show the radial profile of a
star on the same WFPC2 frame in order to illustrate the PSF.}
\label{glob_profile}
\end{figure}

Figure \ref{glob_profile} shows the radial light profile of F8D1-GC1,
along with that of a star on the same frame, and two King (1966) model
profile fits.  As regards the structural parameters given on
Fig.\ \ref{glob_profile}, we note that at the distance of F8D1, log
r$_{\rm core}$ = --0.8 corresponds to r$_{\rm core}$ $\approx$ 3 pc.
The combination of central concentration c = log (r$_{\rm tidal}$ /
r$_{\rm core}$ ) $\gtrsim$ 1.5, r$_{\rm core}$ $\approx$ 3 pc, and
dereddened central surface brightness $\mu_{\rm Vo}$(0) $\sim$ 20.2 is
not common among the Galactic globular clusters.  However, some near
analogs do exist in the compilation of Trager et al.\ (1993).  For
example, NGC~4590 has (c = 1.6, r$_{\rm core}$ = 1.9 pc, and $\mu_{\rm
Vo}$(0) = 18.7) while NGC~2419 has (1.7, 8.4, and 19.6), respectively.
Undoubtedly better estimates (or limits) on the structural parameters
of F8D1-GC1 could be determined by fitting King (1966) models convolved
with the stellar PSF, but such a task is beyond the scope of this
paper.

In summary, we have found one good candidate globular cluster in F8D1,
which is about what we would expect given the relatively high
luminosity of F8D1 and the lack of coverage of the entire galaxy with
the HST images. None were found in BK5N, again as we expected given the
low luminosity of this galaxy.

\section{Discussion} \label{discussion}

We now compare these two dwarf ellipticals with others from the Local
Group that are equally well-studied.  BK5N's luminosity and surface
brightness, as well as its mean metallicity and metallicity dispersion,
place it well within the realm of the Local Group dwarfs: it is very
similar in those respects to the Sculptor galaxy.  The extreme size of
F8D1, however, is not known in the Local Group for galaxies of its
surface brightness.  Its effective radius is in fact larger than any of
the 15 known Local Group dwarf ellipticals (including NGC 205), by at
least a factor of 3, and indeed is larger than that of the SMC and only
15\% smaller than that of the LMC (Bothun \& Thompson 1988).  The
radius of the recently discovered Sgr galaxy appears to be large, but
an accurate number is not yet at hand (Ibata et al.\ 1997).  Its large
ellipticity (e $\approx$ 0.7) and small Galactocentric distance (16
kpc) suggest that the structure of Sgr has been modified by tidal
interactions with the Galaxy.

\begin{figure}[p]
\plotfiddle{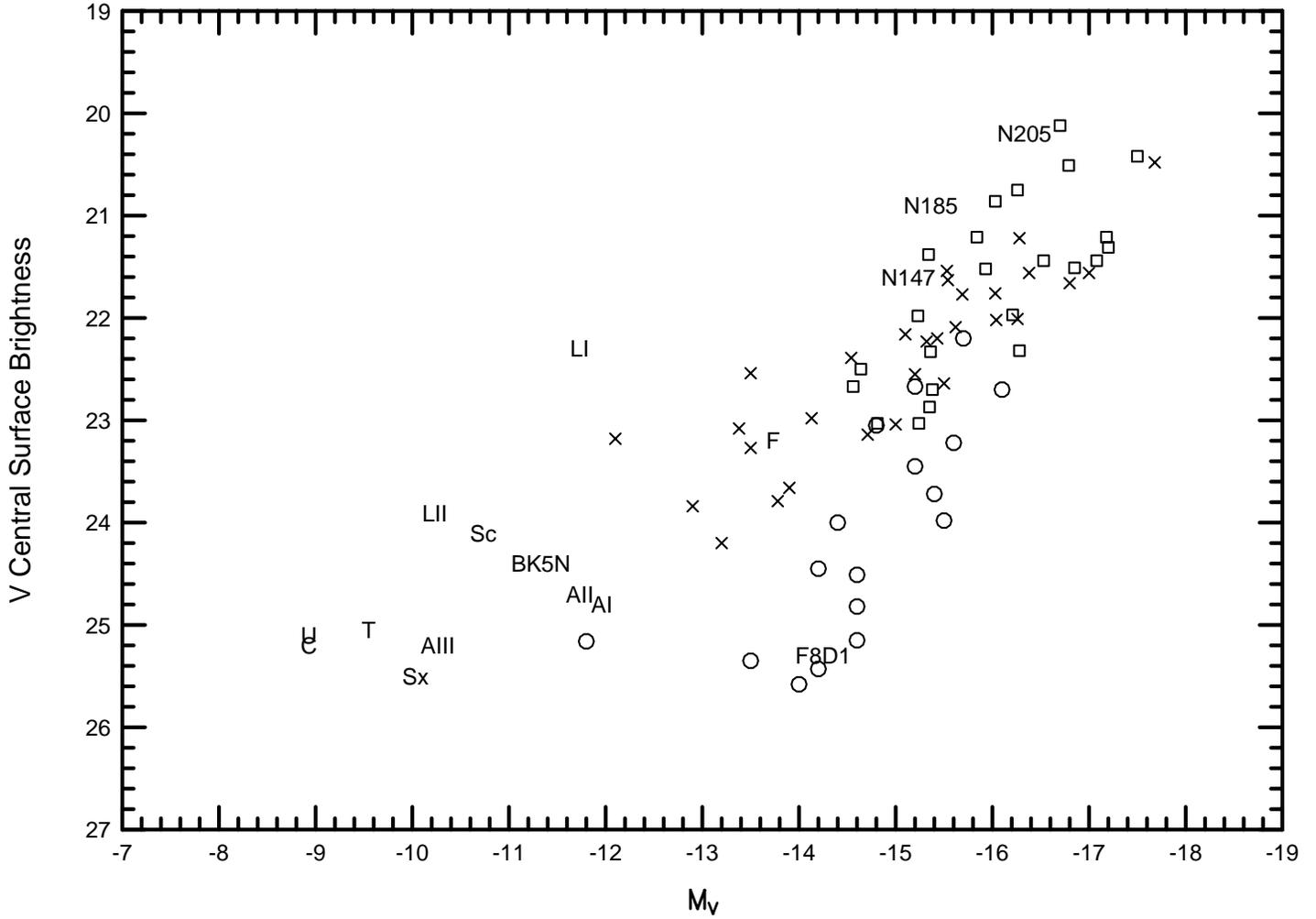}{4.0in}{0.}{78.}{78.}{-283.}{0.}
\caption[]{Central surface brightness -- luminosity relation for
dwarf ellipticals (in $V$). Crosses and open squares represent Virgo and
Fornax dE's; circles represent the large, low surface brightness galaxies
found in Virgo by Impey et al.\ (1988); Local Group dE's are shown
with abbreviations for their names; and the M81 group dwarfs, BK5N
and F8D1, are shown with their names.}
\label{sb_lumin_fig}
\end{figure}

Previously, galaxies like F8D1 have been found in the more numerous
collections of dE's that exist in the large galaxy clusters such as
Virgo and Fornax, as studied by Sandage \& Binggeli (1984), Impey et
al.\ (1988) and Bothun et al.\ (1991).  But they are rare even there.
While Fig.\ \ref{sb_lumin_fig} showed the location in the central
surface brightness--luminosity diagram of the large, low surface
brightness galaxies found by Impey et al., the density of points does
not accurately represent the frequency of occurrence of the galaxies.
As stated above, galaxies as large as F8D1 with central surface
brightnesses of V$_{\rm o} \sim 25.5$ represent less than 1\% of the
total number of galaxies in Virgo and Fornax at that surface
brightness.  Clearly, some physical process during formation drives
most of these galaxies to smaller radii in the mean than we see in
F8D1.  Explaining how such large galaxies come about must involve the
role of supernovae or stellar winds, and would appear to be a central
problem for the formation of dwarf galaxies.

\begin{figure}[p]
\plotfiddle{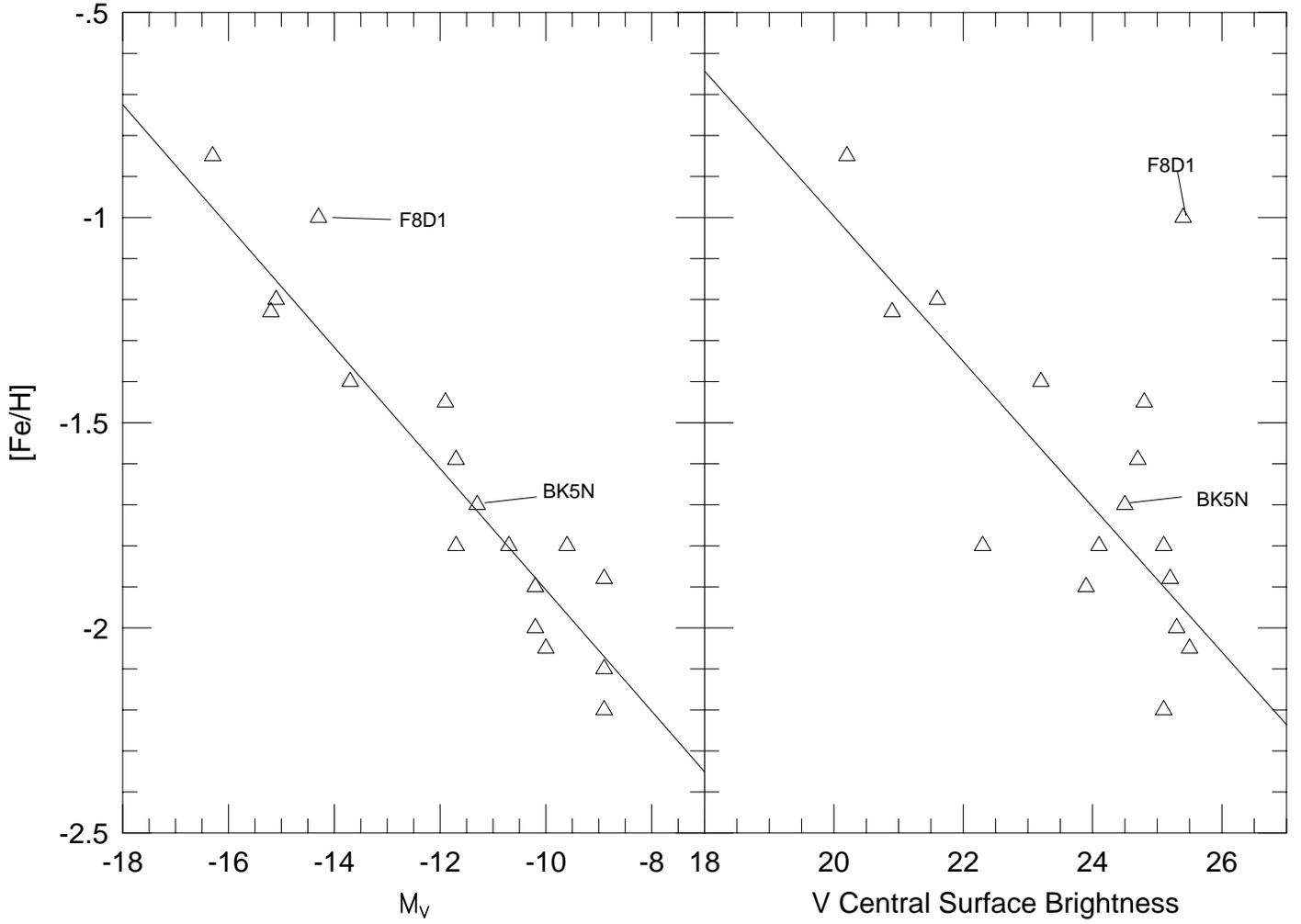}{4.0in}{0.}{78.}{78.}{-283.}{0.}
\caption[]{(a) Luminosity -- metallicity relation for Local Group dwarf
ellipticals, with the two M81 dwarfs plotted and labelled.  (b) Central
surface brightness -- metallicity relation for the same galaxies.  The
line in each panel is a least squares fit to the Local Group data.}
\label{lumin_metal_fig}
\end{figure}

An additional diagnostic relation for dE's is the relation between
luminosity and metallicity (L--Z, here we assume that
[Fe/H] measures Z), wherein the lower luminosity
galaxies have correspondingly lower metallicities. Because lower
luminosity galaxies also tend to have lower surface brightness, there
is as well a general trend of metallicity with central surface
brightness (e.g., Caldwell et al.\ 1992). Several authors have
suggested that indeed it is the local mass density that determines the
mean metallicity of a stellar population, rather than the total system
mass (Phillipps et al.\ 1990; Franx \& Illingworth 1990, the former
commenting on dE's, the latter on E's).  The galaxy F8D1 then provides
a crucial test of this idea, for while it has a very low surface
brightness, it is at the same time luminous, for a dwarf.  Figure
\ref{lumin_metal_fig} shows the relation between L and Z for Local
Group dE's and the two M81 dwarfs for which we have new metallicities,
as well as the relation between central surface brightness V$_{\rm 0}$
and Z (values from Caldwell et
al.\ 1992 and Armandroff et al.\ 1993, updated for the recent results
of Da Costa 1994 and Mateo et al.\ 1993 for Carina, Da Costa et
al.\ 1996 for And I, Lee et al.\ 1993c and Demers et al.\ 1994 for Leo
I, Lee et al.\ 1993b for NGC 185, Konig et al.\ 1993 for And II, and
Saviane et al.\ 1996 for Tucana).

BK5N, as expected, has a metallicity in accord with the relations
defined by the Local Group galaxies, as a function of both luminosity
and surface brightness.  Once again, BK5N is quite similar in all
respects to dwarf ellipticals in the Local Group. Thus, in spite of the
different environments of the M81 group and the Local Group, dwarf
galaxies of very similar properties are present in each.

F8D1 clearly falls near the L--Z relation but deviates significantly
from the V$_{\rm 0}$--Z relation.  The parameters for the recently
discovered Sgr dwarf would seem to place it off of the V$_{\rm 0}$--Z
relation as well, for it has a faint V$_{\rm 0}$ (Ibata et al.\ 1997,
Mateo et al.\ 1995), and yet a relatively high metallicity ([Fe/H]
$\sim -1$ or higher; Mateo et al.\ 1995; Sarajedini \& Layden 1995;
Whitelock et al.\ 1996). However, the properties of Sgr have possibly
been modified by interactions with the Galaxy (Velazquez \& White 1995;
Johnston et al.\ 1995).  For this reason we have excluded Sgr from our
presentation of the L--Z and V$_{\rm 0}$--Z relations.  Given its near
circular isophotes, we have no reason to suspect that F8D1 has been
significantly tidally perturbed by another galaxy, and thus no reason
to question the low central surface brightness of F8D1 nor its total
luminosity.  Thus the location of F8D1 in Fig.\ \ref{lumin_metal_fig}
indicates that the total baryonic mass of a dwarf elliptical dictates
the mean stellar metallicity attained, rather than the baryonic surface
density.

Bender et al.\ (1993) showed that velocity dispersion ($\sigma$, 
meaning central well
potential depth) was well correlated with the spectral index Mg$_2$ for
all E's, giants and dwarfs alike.  Likewise, Peterson \& Caldwell
(1993) found that [Fe/H] and $\sigma$  were correlated for a small
sample of dE's alone.  Using the  $\sigma$--Mg$_2$ relation, Bender
et al.\ could then show from scaling relations that Mg$_2 \propto$
log(M$^2 \rho)$ , where M is the total mass of the system (baryonic and
dark matter), and $\rho$ is the average baryonic density. So that to
the extent that Mg$_2$ measures metallicity, this relation says that
the total mass is indeed the dominant predictor of metallicity, but
that density is important as well.  If density was important in
determining metallicity for dwarfs as well, we might expect to see a
bivariate relation between Z, L and V$_{\rm 0}$, for instance, which
would point to the more fundamental relation between Z and
$\sigma$. (The velocity dispersion for F8D1 would clearly be
desirable in this analysis, but obtaining it is likely to be
difficult).  The data on dwarfs is limited, but it shows that the
addition of a surface brightness term in the fit of L--Z does not
improve the quality of the fit. In particular, the residual for F8D1
from the mean L--Z relation is no larger than for any of the other
galaxies which have luminosities more typical for their surface
brightnesses, so from the present data it appears that L alone predicts
the metallicity for dE's.

There appears to be no additional relation other than the L--Z and
L--V$_{\rm 0}$ relations that would help us to understand the formation
of dwarfs.  A possible relation between L and the Z dispersion, indicated
by the high metallicity spread on the giant branches of Fornax and F8D1
and the low spread of Carina, is ruled out by the high metallicity 
spread of the
low-luminosity Draco (Lehnert et al.\ 1992).  Still, it would appear
that our additional information should help in expanding the recent
theoretical models of dwarf galaxy formation (Dekel \& Silk 1986;
Yoshii \& Arimoto 1987; Navarro et al.\ 1996) to include the large, low
surface brightness galaxies.  In particular, these galaxies may help in
defining the role that dark matter halos play in helping to retain gas
that otherwise may be lost during star formation episodes in dwarfs.

\vskip 0.3truein

Support for this work was provided by NASA through grant number GO-05898
from the Space Telescope Science Institute, which is operated by AURA,
Inc., under NASA contract NAS5-26555.

\clearpage
\begin{figure}[p]
\caption[]{Ground-based $V$-band images of the two M81-group dwarfs
from the FLWO 1.2-m telescope.  Top: F8D1; bottom: BK5N. North is
at the top and east to the left.  An indication of scale is provided
for both panels,
the physical scale quoted assumes a distance of 3.6 Mpc.}
\label{combined_pic}
\end{figure}
\clearpage 

\clearpage

\begin{figure}[p]
\caption{A mosaic of two images of F8D1, taken at the MDM 2.4m telescope,
and showing resolution into stars.  In particular, the asymmetrical
northern part of the galaxy is also resolved into stars.
The images were taken at different
times under different seeing conditions, thus although the FWHM of the
better seeing frame was degraded to match the other, there remains a distinct
difference in the two frames.}
\label{mdm_pic}
\end{figure}

\clearpage

\clearpage

\begin{figure}[p]
\caption[]{WFPC2 footprints on the ground based images of BK5N and F8D1.
The green outlines are for the first visits, red for the second.}
\label{combined_overlay}
\end{figure}

\begin{figure}[p]
\caption[]{PC image of the M81-group dwarf BK5N, made from the
combination of eight 1900 second exposures through the F814W filter.
North is indicated by the direction of the arrow and East by the line.
Both indicators are 2 arcsec in length.}
\label{bk5n_wfpc2}
\end{figure}

\begin{figure}[p]
\caption[]{WFPC2 image of the M81-group dwarf F8D1, made from the
combination of six 1900 second exposures through the F814W filter.
North is indicated by the direction of the arrow and East by the line.
Both indicators are 10 arcsec in length.}
\label{f8d1_wfpc2}
\end{figure}
\clearpage

\clearpage

\begin{deluxetable}{lrr}  
\tablecaption{Photometric Data for BK5N and F8D1}
\tablewidth{4.0in}
\tablecolumns{3}
\tablehead{
\colhead{} & \colhead{BK5N}& \colhead{F8D1}   
} 
\startdata
RA (2000)  &10:04:41.1&9:44:47.1\nl
Dec (2000) &68:15:22~~~&67:26:19~~~\nl
V$_{\rm tot}$ &$ 16.72 \pm 0.25 $&$13.85 \pm 0.25$\nl
M$_{\rm V}$ & --11.33 & --14.25 \nl
B--V & $0.71 \pm 0.05 $&$0.75 \pm 0.12$   \nl
$e$ &0.4 &0.0 \nl
V$_{\rm 0}$ &$ 24.5 \pm 0.2 $&$25.4 \pm 0.2$\nl
S$_{\rm 0}$ &$ 24.92 \pm 0.08$ &$ 25.41 \pm 0.05$\nl
$r_{\rm 0}$ ~(\arcsc)  &$ 24 \pm 1 $&$ 106 \pm 5$\nl
n& $1.81 \pm 0.09 $ &$ 1.45 \pm 0.08$ \nl
R$_{\rm eff}$~(\arcsc) &20 &127 \nl
R$_{\rm eff}$~(pc) & 370 & 2450 \nl
R$_{\rm c}$~(\arcsc) &$ 16 \pm 1$&$82 \pm 2$ \nl
E(B-V) &0.048 & 0.032 \nl
(m--M)$_0$&$27.9 \pm 0.15$&$28.0 \pm 0.1$\nl
\enddata
\noindent
\tablenotetext{}{V$_{\rm 0}$ is the observed central surface brightness
in V.  S$_{\rm 0}$, $r_{\rm 0}$, and n are parameters for the Sersic
profile fit.  R$_{\rm eff}$ is the effective radius.  R$_{\rm c}$ is
the core radius (radius at which surface brightness falls to half the
central value).}
\label{phot_table}
\end{deluxetable}

\begin{deluxetable}{llcclcc}
\tablecolumns{7}
\tablecaption{HST Observing Log for BK5N and F8D1}
\tablewidth{0pt}
\tablehead{
\colhead{} &\multicolumn{3}{c}{Visit 1} &\multicolumn{3}{c}{Visit 2}\\
\colhead{} &\colhead{date} &\colhead{F555W}&\colhead {F814W}& \colhead{date} &\colhead {F555W}&\colhead {F814W}
} 
\startdata
BK5N & 1996Feb22 &$3\times 1800$s & $6\times1900$s  &1996Jun9-11 &$6\times 1800$s & $8\times1900$s  \nl
F8D1 & 1996Feb23 &$5\times 1800$s &$6\times1900$s &1996Apr17 &$5\times 1800$s  & $8\times1900$s  \nl
\enddata
\label{obs_log_table}
\end{deluxetable}

\begin{deluxetable}{lrrcccc}
\tablecolumns{7}
\tablecaption{Uncertainties in the Photometry}
\tablehead{
\colhead{galaxy}&\colhead{$I_{\rm midpt}$}&\colhead{N}&
\colhead{$<\sigma_{\rm photon}>_I$}&
\colhead{$<\sigma_{\rm repeat}>_I$}&
\colhead{$<\sigma_{\rm photon}>_{V-I}$}&
\colhead{$<\sigma_{\rm repeat}>_{V-I}$}}
\startdata
F8D1 & 23.00& 16 & 0.011 & 0.019 & 0.037 & 0.038 \nl
F8D1 & 23.50& 54 & 0.016 & 0.029 & 0.048 & 0.049 \nl
F8D1 & 24.00&174 & 0.024 & 0.037 & 0.061 & 0.062 \nl
F8D1 & 24.50&399 & 0.034 & 0.040 & 0.080 & 0.081 \nl
F8D1 & 25.00&570 & 0.052 & 0.062 & 0.111 & 0.117 \nl
F8D1 & 25.50&431 & 0.077 & 0.077 & 0.151 & 0.155 \nl
F8D1 & 26.00&187 & 0.110 & 0.115 & 0.203 & 0.210 \nl
 \nl
BK5N & 23.50&  8 & 0.014 & 0.023 & 0.033 & 0.045 \nl
BK5N & 24.00& 32 & 0.020 & 0.027 & 0.040 & 0.040 \nl
BK5N & 24.50& 78 & 0.027 & 0.032 & 0.053 & 0.062 \nl
BK5N & 25.00& 90 & 0.042 & 0.044 & 0.076 & 0.080 \nl
BK5N & 25.50& 86 & 0.062 & 0.066 & 0.107 & 0.105 \nl
BK5N & 26.00& 47 & 0.085 & 0.092 & 0.143 & 0.146 \nl
\enddata
\label{error_tab}
\end{deluxetable}

\begin{deluxetable}{rllrrrrr}
\tablecolumns{8}
\tablecaption{Variable Stars in BK5N \& F8D1}
\tablewidth{0pt}
\tablehead{
\colhead{ID}& \colhead{$<$I$>$}& \colhead{$<$V--I$>$}& \colhead{$\Delta $V}&\colhead{$\Delta $I}& \colhead{$\frac{\Delta \rm V}{\sigma}$}&\colhead{$\frac{\Delta \rm I}{\sigma}$}& \colhead{$\frac{\Delta \rm{(V-I)}}{\sigma}$}
}
\startdata
\multicolumn{8}{c}{BK5N Variables~~~($\Delta $T=109 days)}\\
1300-1 &  $ 23.656 \pm 0.014$& $1.687 \pm 0.031$&$   -0.37$&$  -0.12$&$-6.60$&$ -4.50$&$  -3.95$\nl
1803-1 &  $ 24.054 \pm 0.022$& $1.457 \pm 0.040$&$   -0.60$&$  -0.44$&$-8.84$&$-10.18$&$  -1.93$\nl
1839-1 &  $ 23.440 \pm 0.012$& $2.202 \pm 0.036$&$    0.92$&$   0.52$&$13.61$&$ 21.79$&$   4.02$\nl
1943-1 &  $ 24.720 \pm 0.032$& $1.193 \pm 0.054$&$    0.50$&$   0.41$&$5.82 $&$ 6.84 $&$   0.66$\nl
\tableline
\multicolumn{8}{c}{BK5N Probable Variables}\\
 519-2 &  $ 23.406 \pm 0.031$& $2.047 \pm 0.041$&$    0.33$&$   0.08$&$ 4.23$&$  3.00$&$   2.73$\nl
1069-1 &  $ 22.598 \pm 0.007$& $1.712 \pm 0.015$&$   -0.17$&$  -0.01$&$-6.57$&$ -1.07$&$  -5.17$\nl
1170-1 &  $ 24.187 \pm 0.020$& $1.541 \pm 0.042$&$    0.21$&$   0.18$&$ 2.87$&$  4.72$&$   0.26$\nl
1353-1 &  $ 25.042 \pm 0.043$& $1.170 \pm 0.071$&$   -0.45$&$  -0.40$&$-3.97$&$ -4.70$&$  -0.33$\nl
1372-1 &  $ 23.358 \pm 0.011$& $1.811 \pm 0.027$&$    0.02$&$   0.09$&$ 0.50$&$  4.18$&$  -1.19$\nl
1436-1 &  $ 24.275 \pm 0.022$& $1.343 \pm 0.041$&$   -0.28$&$   0.05$&$-4.08$&$  1.31$&$  -4.18$\nl
1540-1 &  $ 23.770 \pm 0.014$& $2.227 \pm 0.048$&$    0.28$&$   0.23$&$ 3.08$&$  8.17$&$   0.49$\nl
1661-1 &  $ 24.505 \pm 0.027$& $1.430 \pm 0.051$&$    0.08$&$   0.22$&$ 0.90$&$  4.09$&$  -1.33$\nl
\tablebreak 
\multicolumn{8}{c}{F8D1 Variables~~~($\Delta $T=56 days)}\\
 603-4 &  $ 23.868 \pm 0.019$& $1.053 \pm 0.031$&$   -0.87$&$  -0.23$&$-17.4$&$-6.11 $&$  -8.89$\nl
 708-4 &  $ 23.018 \pm 0.010$& $2.151 \pm 0.032$&$    0.37$&$   0.20$&$ 6.20$&$ 10.35$&$   2.40$\nl
 899-2 &  $ 23.841 \pm 0.019$& $2.697 \pm 0.093$&$    0.78$&$   0.42$&$ 4.28$&$ 11.13$&$   1.41$\nl
 978-2 &  $ 23.493 \pm 0.015$& $2.878 \pm 0.084$&$    0.79$&$   0.50$&$ 4.76$&$ 16.63$&$   1.17$\nl
1118-4 &  $ 23.252 \pm 0.012$& $1.934 \pm 0.034$&$    0.27$&$   0.13$&$ 4.23$&$  5.50$&$   1.95$\nl
1201-2 &  $ 23.056 \pm 0.010$& $2.137 \pm 0.033$&$    0.41$&$   0.21$&$ 6.62$&$ 10.55$&$   2.78$\nl
1372-1 &  $ 23.461 \pm 0.026$& $1.828 \pm 0.063$&$   -0.59$&$  -0.28$&$-5.18$&$ -5.35$&$  -2.16$\nl
1360-2 &  $ 23.315 \pm 0.014$& $2.409 \pm 0.052$&$   -0.92$&$  -0.27$&$-9.20$&$ -9.82$&$  -4.47$\nl
1457-1 &  $ 23.593 \pm 0.012$& $2.960 \pm 0.065$&$    0.75$&$   0.34$&$ 5.84$&$ 14.33$&$   2.77$\nl
1515-2 &  $ 23.566 \pm 0.016$& $2.178 \pm 0.053$&$    0.41$&$   0.17$&$ 4.09$&$  5.53$&$   1.97$\nl
1535-2 &  $ 22.961 \pm 0.010$& $2.454 \pm 0.040$&$   -0.36$&$  -0.13$&$-4.73$&$ -6.85$&$  -2.64$\nl
1731-3 &  $ 23.878 \pm 0.015$& $4.317 \pm 0.235$&$   -2.84$&$  -0.52$&$-6.03$&$-17.27$&$  -1.28$\nl
1796-3 &  $ 23.490 \pm 0.013$& $2.340 \pm 0.053$&$   -1.00$&$  -0.30$&$-9.88$&$-11.76$&$  -5.46$\nl
1808-3 &  $ 23.607 \pm 0.016$& $2.267 \pm 0.056$&$   -0.54$&$  -0.56$&$-5.02$&$-17.50$&$   0.13$\nl
1977-2 &  $ 25.602 \pm 0.088$& $1.027 \pm 0.136$&$    1.07$&$   1.00$&$ 5.16$&$ 5.71 $&$   0.16$\nl
2227-2 &  $ 23.596 \pm 0.015$& $1.746 \pm 0.039$&$    0.56$&$   0.27$&$ 7.87$&$  9.10$&$   3.43$\nl
2482-2 &  $ 23.060 \pm 0.010$& $2.423 \pm 0.038$&$    0.36$&$   0.24$&$ 4.91$&$ 12.10$&$   1.46$\nl
2918-2 &  $ 22.664 \pm 0.007$& $2.484 \pm 0.030$&$   -0.41$&$  -0.16$&$-7.19$&$ -8.35$&$  -4.07$\nl
\tableline
\multicolumn{8}{c}{F8D1 Probable Variables}\\
 616-4 &  $24.452 \pm 0.032$&$ 1.633 \pm 0.071  $&$  0.23$&$   0.35$&$ 1.84$&$   5.51$&$   -0.79   $\nl
 667-4 &  $23.980 \pm 0.022$&$ 1.812 \pm 0.056  $&$  0.24$&$   0.26$&$ 2.36$&$   6.06$&$   -0.21   $\nl
 686-4 &  $24.016 \pm 0.022$&$ 1.787 \pm 0.056  $&$  0.03$&$   0.22$&$ 0.28$&$   5.18$&$   -1.64   $\nl
 723-3 &  $23.791 \pm 0.018$&$ 2.736 \pm 0.091  $&$  0.24$&$   0.15$&$ 1.38$&$   4.22$&$    0.46   $\nl
 796-3 &  $23.776 \pm 0.017$&$ 3.006 \pm 0.111  $&$  0.45$&$   0.26$&$ 2.05$&$   7.85$&$    0.67   $\nl
 815-4 &  $19.855 \pm 0.001$&$ 2.615 \pm 0.004  $&$  0.07$&$   0.17$&$ 3.75$&$   8.50$&$  -10.62   $\nl
 858-4 &  $23.466 \pm 0.015$&$ 2.503 \pm 0.061  $&$  0.26$&$   0.17$&$ 2.27$&$   5.96$&$    0.67   $\nl
 872-3 &  $22.982 \pm 0.009$&$ 2.709 \pm 0.047  $&$  0.12$&$   0.12$&$ 1.30$&$   6.45$&$   -0.09   $\nl
 976-4 &  $23.384 \pm 0.013$&$ 2.028 \pm 0.039  $&$  0.11$&$   0.12$&$ 1.51$&$   4.65$&$   -0.10   $\nl
 999-4 &  $22.857 \pm 0.009$&$ 1.917 \pm 0.025  $&$  0.05$&$   0.13$&$ 1.15$&$   6.75$&$   -1.64   $\nl
1000-4 &  $23.306 \pm 0.012$&$ 2.119 \pm 0.037  $&$  0.11$&$   0.13$&$ 1.61$&$   5.79$&$   -0.33   $\nl
1069-2 &  $23.610 \pm 0.016$&$ 2.149 \pm 0.051  $&$  0.33$&$   0.13$&$ 3.46$&$   4.09$&$    1.81   $\nl
1090-2 &  $23.502 \pm 0.015$&$ 2.531 \pm 0.063  $&$  0.20$&$   0.15$&$ 1.66$&$   5.20$&$    0.34   $\nl
1095-2 &  $25.499 \pm 0.081$&$ 1.168 \pm 0.136  $&$  0.09$&$   0.75$&$ 0.44$&$   4.66$&$   -2.02   $\nl
1149-2 &  $25.371 \pm 0.075$&$ 1.088 \pm 0.121  $&$ -0.47$&$  -0.63$&$ 2.46$&$  -4.26$&$    0.60   $\nl
1168-2 &  $23.149 \pm 0.010$&$ 2.457 \pm 0.044  $&$ -0.31$&$  -0.20$&$ 3.70$&$ -10.30$&$   -1.18   $\nl
1329-4 &  $23.472 \pm 0.014$&$ 2.166 \pm 0.047  $&$  0.17$&$   0.13$&$ 1.93$&$   4.82$&$    0.38   $\nl
1337-2 &  $25.344 \pm 0.089$&$ 1.238 \pm 0.154  $&$ -0.23$&$  -0.72$&$ 0.91$&$  -4.05$&$    1.36   $\nl
1382-4 &  $23.677 \pm 0.014$&$ 3.857 \pm 0.179  $&$ -0.12$&$   0.17$&$ 0.35$&$   6.03$&$   -0.67   $\nl
1399-3 &  $23.430 \pm 0.013$&$ 3.007 \pm 0.077  $&$  0.02$&$  -0.20$&$ 0.14$&$  -8.03$&$    1.36   $\nl
1411-2 &  $23.420 \pm 0.013$&$ 2.749 \pm 0.069  $&$  0.01$&$  -0.14$&$ 0.08$&$  -5.38$&$    1.01   $\nl
1432-3 &  $23.539 \pm 0.016$&$ 2.391 \pm 0.058  $&$  0.17$&$   0.13$&$ 1.59$&$   4.31$&$    0.32   $\nl
1445-2 &  $24.464 \pm 0.035$&$ 1.454 \pm 0.071  $&$ -0.49$&$  -0.33$&$ 3.97$&$  -4.80$&$   -0.99   $\nl
1463-1 &  $23.789 \pm 0.015$&$ 2.222 \pm 0.045  $&$ -0.32$&$  -0.16$&$-3.85$&$  -5.27$&$   -1.69   $\nl
1615-1 &  $23.380 \pm 0.010$&$ 1.977 \pm 0.027  $&$  0.07$&$   0.14$&$ 1.36$&$   7.05$&$   -1.31   $\nl
1621-1 &  $23.715 \pm 0.013$&$ 2.792 \pm 0.060  $&$  0.16$&$   0.11$&$ 1.38$&$   4.23$&$    0.46   $\nl
1629-3 &  $23.681 \pm 0.015$&$ 3.195 \pm 0.109  $&$ -0.51$&$  -0.16$&$ 2.40$&$  -5.60$&$   -1.28   $\nl
1657-2 &  $23.762 \pm 0.019$&$ 2.427 \pm 0.075  $&$ -0.56$&$  -0.17$&$ 3.87$&$  -4.57$&$   -2.18   $\nl
1689-1 &  $23.538 \pm 0.012$&$ 2.823 \pm 0.055  $&$  0.17$&$   0.10$&$ 1.56$&$   4.08$&$    0.61   $\nl
1835-2 &  $24.359 \pm 0.028$&$ 1.693 \pm 0.069  $&$  0.16$&$   0.22$&$ 1.27$&$   4.01$&$   -0.43   $\nl
1845-3 &  $23.481 \pm 0.014$&$ 2.568 \pm 0.058  $&$  0.18$&$   0.27$&$ 1.67$&$   9.75$&$   -0.69   $\nl
2078-2 &  $23.398 \pm 0.019$&$ 2.258 \pm 0.062  $&$  0.25$&$   0.41$&$ 2.12$&$  11.00$&$   -1.21   $\nl
2088-3 &  $23.530 \pm 0.013$&$ 2.446 \pm 0.056  $&$  0.34$&$   0.21$&$ 3.23$&$   8.07$&$    1.12   $\nl
2110-2 &  $23.794 \pm 0.018$&$ 2.915 \pm 0.109  $&$ -0.61$&$  -0.23$&$ 2.86$&$  -6.52$&$   -1.38   $\nl
2218-3 &  $23.554 \pm 0.014$&$ 2.588 \pm 0.066  $&$ -0.24$&$  -0.12$&$ 1.85$&$  -4.46$&$   -0.81   $\nl
2467-3 &  $22.959 \pm 0.009$&$ 1.709 \pm 0.022  $&$  0.03$&$   0.09$&$ 0.90$&$   4.55$&$   -1.24   $\nl
2511-2 &  $24.988 \pm 0.049$&$ 1.599 \pm 0.112  $&$  0.17$&$  -0.45$&$ 0.86$&$  -4.60$&$    2.46   $\nl
2516-2 &  $24.106 \pm 0.024$&$ 2.062 \pm 0.073  $&$ -0.45$&$  -0.20$&$ 3.30$&$  -4.25$&$   -1.53   $\nl
\enddata
\label{variables_tab}
\end{deluxetable}

\clearpage
\begin{deluxetable}{llrrrr}
\tablecolumns{6}
\tablecaption{Data for Candidate Globular Cluster in F8D1}
\tablewidth{0pt}
\tablehead{
\colhead{RA (2000)}&\colhead{Dec (2000)}&\colhead{V}&\colhead{M$_{\rm V}$}& \colhead{V--I}
}
\startdata
9:44:39.2 & 67:26:06 & $21.68 \pm 0.05$ & $-6.2$ & $0.76 \pm 0.01 $ \nl
\enddata
\label{glob_tab}
\end{deluxetable}

\end{document}